\newif\ifsubmission
\submissiontrue 

\newif\ifanonymous
\anonymousfalse

\newif\iffull
\fullfalse

\documentclass[letterpaper,twocolumn,10pt]{article}
\usepackage{usenix2019}
\usepackage{endnotes}
\usepackage{cite}
\usepackage{amsmath,amssymb,amsfonts}

\usepackage{algorithm}
\usepackage{algpseudocode}
\usepackage{pseudocode}
\usepackage{multicol}

\usepackage{graphicx}
\usepackage{textcomp}
\usepackage{xcolor}
\def\BibTeX{{\rm B\kern-.05em{\sc i\kern-.025em b}\kern-.08em
    T\kern-.1667em\lower.7ex\hbox{E}\kern-.125emX}}

\usepackage{colortbl}
\usepackage{endnotes}
\usepackage{booktabs}
\usepackage{graphicx}
\usepackage{amsmath}
\usepackage{amsfonts}
\usepackage{amssymb}
\usepackage{multirow}
\usepackage{url}

\usepackage{tikz}
\usepackage{tikzscale}
\usepackage{cancel}
\usepackage{todonotes}
\usepackage{tipa}
\usepackage{color}
\usepackage{bm}
\usepackage{subcaption}
\usepackage{enumitem}
\usepackage{listings}
\usepackage{array}
\usepackage{footnote}
\usepackage{numprint}
\usepackage{wrapfig}
\usepackage{ragged2e}
\usepackage{lipsum}
\usepackage{cleveref}
\usepackage[binary-units=true, per-mode=symbol, per-symbol=/]{siunitx}
\usepackage[multiple]{footmisc}
\usepackage{csquotes}
\usepackage{tabularx}
\usepackage{pifont}
\usepackage{xcolor}
\usepackage{courier}
\usepackage{tcolorbox}
\usepackage{mathtools}
\usepackage{stfloats}
\usepackage{supertabular}
\usepackage{tipa}
\usepackage[square,sort,comma,numbers]{./natbib}

\usepackage{etoolbox}
\patchcmd{\thebibliography}
  {\settowidth}
  {\setlength{\itemsep}{0pt plus 0pt}\settowidth}
  {}{}
\apptocmd{\thebibliography}
  {\small}
  {}{}
\newcommand{\cmark}{\ding{51}}
\newcommand{\xmark}{\ding{55}}

\newcolumntype{C}[1]{>{\centering\arraybackslash}p{#1}}
\makesavenoteenv{tabular}
\npthousandsep{,}\npthousandthpartsep{}\npdecimalsign{.}
\graphicspath{{figs/}} 
\newcolumntype{R}[1]{>{\raggedleft\let\newline\\\arraybackslash\hspace{0pt}}m{#1}}

\makeatletter

\crefformat{section}{\S#2#1#3} 
\crefformat{subsection}{\S#2#1#3}
\crefformat{subsubsection}{\S#2#1#3}

\ifsubmission
  \newcommand{\TODO}[1]{}
  
  \newcommand{\rebut}[1]{#1}
\else 
  
  \newcommand{\TODO}[1]{\textcolor{red}{TODO: #1}}
  
  \newcommand{\rebut}[1]{\textcolor{red}{#1}}
\fi

\newcommand{\mli}[1]{\mathit{#1}}
\DeclareMathOperator*{\argmax}{arg\,max}

\newcommand{\remove}[1]{}

\newcommand{\sect}[1]{Section~\ref{#1}}
\newcommand{\fig}[1]{Figure~\ref{#1}}
\newcommand{\tab}[1]{Table~\ref{#1}}

\newcommand{\app}[1]{Appendix~\ref{#1}}
\newcommand{\sys}[1]{\textsc{Xonn}}
\newcommand{\gazelle}[1]{Gazelle}

\newcommand\mc[1]{\multicolumn{1}{|c|}{#1}} 

\newcommand{\mypm}{\mathbin{\mathpalette\@mypm\relax}}
\newcommand{\@mypm}[2]{\ooalign{%
  \raisebox{.1\height}{$#1+$}\cr
  \smash{\raisebox{-.6\height}{$#1-$}}\cr}}

\newcommand{\lname}[1]{\textsf{#1}}

\DeclarePairedDelimiter\ceil{\lceil}{\rceil}

\makeatother

\usepackage[absolute,overlay]{textpos}

\begin{document}
\setlength{\abovedisplayskip}{3pt}
\setlength{\belowdisplayskip}{3pt}

\begin{textblock*}{8cm}(1.5cm,1.5cm) 
   To appear in USENIX Security 2019
\end{textblock*}

\title{\vspace{-1cm}\sys{}: \texttt{XNOR}-based Oblivious Deep Neural Network Inference \vspace{-0.00cm}
}
\ifanonymous
    \author{}
\else
    \author{
    {\rm M. Sadegh Riazi}\\
    UC San Diego
    \and
    {\rm Mohammad Samragh}\\
    UC San Diego
    \and
    {\rm Hao Chen}\\
    Microsoft Research
    \and
    {\rm Kim Laine}\\
    Microsoft Research
    \and
    {\rm Kristin Lauter}\\
    Microsoft Research
    \and
    {\rm Farinaz Koushanfar}\\
    UC San Diego
    \\[-3.0cm]
}
\fi


\maketitle


\subsection*{Abstract}
Advancements in deep learning enable cloud servers to provide inference-as-a-service for clients. 
In this scenario, clients send their raw data to the server to run the deep learning model and send back the results. 
One standing challenge in this setting is to ensure the privacy of the clients' sensitive data.  
Oblivious inference is the task of running the neural network on the client's input without disclosing the input or the result to the server. 
This paper introduces \sys{}
(pronounced /\textipa{z2n}/), a novel end-to-end framework based on Yao's Garbled Circuits (GC) protocol, that provides a paradigm shift in the conceptual and practical realization of oblivious inference. 
In \sys{}, the costly matrix-multiplication operations of the deep learning model are replaced with \texttt{XNOR} operations that are essentially free in GC. 
We further provide a novel algorithm that customizes the neural network such that the runtime of the GC protocol is minimized without sacrificing the inference accuracy. 

We design a user-friendly high-level API for \sys{}, allowing expression of the deep learning model architecture in an unprecedented level of abstraction.
We further provide a compiler to translate the model description from high-level Python (i.e., Keras) to that of \sys{}. 
Extensive proof-of-concept evaluation on various neural network architectures demonstrates that \sys{} outperforms prior art such as Gazelle (USENIX Security'18) by up to $7\times$, MiniONN (ACM CCS'17) by $93\times$, and SecureML (IEEE S\&P'17) by $37\times$. State-of-the-art frameworks require one round of interaction between the client and the server for each layer of the neural network, whereas, \sys{} requires a {\it constant} round of interactions for {\it any} number of layers in the model. 
\sys{} is first to perform oblivious inference on Fitnet architectures with up to $21$ layers, suggesting a new level of scalability compared with state-of-the-art.
Moreover, we evaluate \sys{} on four datasets to perform privacy-preserving medical diagnosis. The datasets include breast cancer, diabetes, liver disease, and Malaria.


\section{Introduction}\label{sec:intro}
The advent of big data and striking recent progress in artificial intelligence are fueling the impending industrial automation revolution. In particular, Deep Learning (DL) \textemdash a method based on learning Deep Neural Networks (DNNs) \textemdash is demonstrating a breakthrough in accuracy. DL models outperform human cognition in a number of critical tasks such as speech and visual recognition, natural language processing, and medical data \rebut{analysis}. Given DL's superior performance, several technology companies are now developing or already providing DL as a service. They train their DL models on a large amount of (often) proprietary data on their own servers; then, an inference API is provided to the users who can send their data to the server and receive the analysis results on their queries.   
The notable shortcoming of this remote inference service is that the inputs are revealed to the cloud server, breaching the privacy of sensitive user data.

Consider a DL model used in a medical task in which a health service provider withholds the prediction model. Patients submit their plaintext medical information to the server, which then uses the sensitive data to provide a medical diagnosis based on inference obtained from its proprietary model.
 A naive solution to ensure patient privacy is to allow the patients to receive the DL model and run it on their own trusted platform. However, this solution is not practical in real-world scenarios because: 
 (i)~The DL model is considered an essential component of the service provider's intellectual property (IP). Companies invest a significant amount of resources and funding to gather the massive datasets and train the DL models; hence, it is important to service providers not to reveal the DL model to ensure their profitability and competitive advantage.
 (ii) The DL model is known to reveal information about the underlying data used for training~\cite{tramer2016stealing}. In the case of medical data, this reveals sensitive information about other patients, violating HIPAA and similar patient health privacy regulations.

{\it Oblivious inference} is the task of running the DL model on the client's input without disclosing the input \rebut{or} the result to the server itself. Several solutions for oblivious inference have been proposed that utilize one or more cryptographic tools such as Homomorphic Encryption (HE)~\cite{brakerski2014efficient,brakerski2014leveled}, Garbled Circuits (GC)~\cite{yao1986generate}, Goldreich-Micali-Wigderson (GMW) protocol~\cite{goldreich1987play}, and Secret Sharing (SS). Each of these cryptographic tools offer their own characteristics and trade-offs. 
For example, one major drawback of HE is its {\it computational complexity}. HE has two main variants: Fully Homomorphic Encryption (FHE)~\cite{brakerski2014efficient} and Partially Homomorphic Encryption (PHE)~\cite{brakerski2014leveled,paillier1999public}. FHE allows computation on encrypted data but is computationally very expensive. PHE has less overhead but only supports a subset of functions or depth-bounded arithmetic circuits. 
The computational complexity drastically increases with the circuit's depth. Moreover, non-polynomial functionalities such as the ReLU activation function in DL cannot be supported. 

GC, on the other hand, can support an arbitrary functionality while \rebut{requiring only} a {\it constant} round of interactions regardless of the depth of the computation. However, it has a high communication cost and a significant overhead for multiplication. More precisely, performing multiplication in GC has quadratic computation and communication complexity with respect to the bit-length of the input operands. It is well-known that the complexity of the contemporary DL methodologies is dominated by matrix-vector multiplications. 
GMW needs less communication than GC but requires many rounds of {\it interactions} between the two parties. 

A standalone SS-based scheme provides a computationally inexpensive multiplication yet requires three or more independent (non-colluding) computing servers, which is a strong assumption.  
Mixed-protocol solutions have been proposed with the aim of utilizing the best characteristics of each of these protocols~\cite{chameleon, secureml, minionn, gazelle}. They require \rebut{secure conversion of} secrets from one protocol to another in the middle of execution. Nevertheless, it has been shown that the cost of secret conversion is paid off in these hybrid solutions. Roughly speaking, \rebut{the number of interactions between server and client (i.e., round complexity) in existing hybrid solutions is} {\it linear} with respect to the depth of the DL model. Since depth is a major contributor to the deep learning accuracy~\cite{szegedy2015going}, scalability of the mixed-protocol solutions with respect to the number of layers remains an unsolved issue for more complex, \rebut{many-layer} networks.

This paper introduces \sys{}, a novel end-to-end framework which provides a paradigm shift in the conceptual and practical realization of privacy-preserving interference on deep neural networks. The existing work has largely focused on the development of customized security protocols while using conventional fixed-point deep learning algorithms. \sys{}, for the first time, suggests leveraging the concept of the Binary Neural Networks (BNNs) in conjunction with the GC protocol. 
In BNNs, the weights and activations are restricted to binary (i.e, $\pm1$) values, substituting the costly multiplications with simple \texttt{XNOR} operations during the inference phase. The \texttt{XNOR} operation is known to be {\it free} in the GC protocol~\cite{freexor}; therefore, performing oblivious inference on BNNs using GC results in the removal of costly multiplications.
Using our approach, we show that oblivious inference on the standard DL benchmarks can be performed with minimal, \rebut{if any}, decrease in the prediction accuracy.

We emphasize that an effective solution for oblivious inference should take into account the deep learning algorithms and optimization methods that can tailor the DL model for the security protocol. Current DL models are designed to run on CPU/GPU platforms where many multiplications can be performed with high throughput, whereas, bit-level operations are very inefficient. In the GC protocol, however, bit-level operations are inexpensive, but multiplications are rather costly. As such, we propose to train deep neural networks that involve many bit-level operations but \textit{no} multiplications in the inference phase; using the idea of learning binary networks, we achieve an average of $21\times$ reduction in the number of gates for the GC protocol.


We perform extensive evaluations on different datasets.
Compared to the \gazelle{}~\cite{gazelle} (the prior best solution) and MiniONN~\cite{minionn} frameworks, we achieve $7\times$ and $93\times$ lower inference latency, respectively.
\sys{} outperforms DeepSecure~\cite{deepsecure} (prior best GC-based framework) by $60\times$ and CryptoNets~\cite{cryptonets}, an HE-based framework, by $1859\times$.
Moreover, our solution renders a {\it constant} round of interactions between the client and the server, which has a significant effect on the performance on oblivious inference in Internet settings. We highlight our contributions as follows:

\begin{itemize}[leftmargin=*]
\vspace{-0.4em}
\item Introduction of \sys{}, the \textit{first} framework for privacy preserving DNN inference with a \textit{constant} round complexity that does not need expensive matrix multiplications. Our solution is the first that can be scalably adapted to ensure security against malicious adversaries.
\item Proposing a novel conditional addition protocol based on Oblivious Transfer (OT)~\cite{rabin2005exchange}, which optimizes the costly computations for the network's input layer. Our protocol is $6\times$ faster than GC and can be of independent interest. We also devise a novel network trimming algorithm to remove neurons from DNNs that minimally contribute to the inference accuracy, further reducing the GC complexity. 
\item Designing a high-level API to readily automate fast adaptation of \sys{}, such that users only input a high-level description of the neural network. We further facilitate the usage of our framework by designing a compiler that translates the network description from Keras to \sys{}. 
\item Proof-of-concept implementation of \sys{} and evaluation on various standard deep learning benchmarks. To demonstrate the scalability of \sys{}, we perform oblivious inference on neural networks with as many \rebut{as $21$ layers} for the first time in the oblivious inference literature.
\end{itemize}

\vspace{-1em}
\section{Preliminaries}\label{sec:prelim}
\vspace{-0.7em}
Throughout this paper, scalars are represented as lowercase letters ($x \in \mathbb{R}$), vectors are represented as bold lowercase letters ($\mathbf{x} \in \mathbb{R}^n$), matrices are denoted as capital letters ($X\in \mathbb{R}^{m\times n}$), and tensors of more than 2 ways are shown using bold capital letters ($\mathbf{X}\in \mathbb{R}^{m \times n \times k}$). Brackets denote element selection and the colon \rebut{symbol} stands for all elements \textemdash $W[i,:]$ represents all values in the $i$-th row of $W$.

\vspace{-0.7em}
\subsection{Deep Neural Networks}\label{ssec:dnn}
The computational flow of a deep neural network is composed of multiple computational layers. The input to each layer is either a vector (i.e., $\mathbf{x} \in \mathbb{R}^n$) or a tensor (i.e., $\mathbf{X} \in \mathbb{R}^{m\times n\times k}$). The output of each layer serves as the input of the \rebut{next} layer. The input of the first layer is the raw data and the output of the last layer represents the network's prediction on the given data (i.e., inference result). In an image classification task, for instance, \rebut{the raw image serves as the input to the first layer} and the output of the last layer is a vector whose elements represent the probability that the image belongs to each category. Below we describe the functionality of neural network layers.





\vspace{0.2em}
{\noindent \bf Linear Layers:}
Linear operations in neural networks are performed in Fully-Connected (\lname{FC}) and Convolution (\lname{CONV}) layers. The vector dot product (\lname{VDP}) between two vectors $\mathbf{x}\in \mathbb{R}^n$ and $\mathbf{w}\in \mathbb{R}^n$ is defined as follows:
\begin{equation}
    \lname{VDP}\ (\mathbf{x}, \mathbf{w}) =\sum_{i=1}^{n} \mathbf{w}[i]\cdot \mathbf{x}[i].
\end{equation}
Both \lname{CONV} and \lname{FC} layers repeat \lname{VDP} computation to generate outputs as we describe next.
A fully connected layer takes a vector $\mathbf{x}\in \mathbb{R}^n$ and generates the output $\mathbf{y} \in \mathbb{R}^m$ using a linear transformation:
\begin{equation}
    \mathbf{y} = W\cdot \mathbf{x} + \mathbf{b},
\end{equation}
where $W \in \mathbb{R}^{m\times n}$ is the weight matrix and $\mathbf{b}\in \mathbb{R}^m$ is a bias vector. More precisely, the $i$-th \rebut{output element is computed as $\mathbf{y[i]}=\lname{VDP}\ (W[i,:],\mathbf{x})+\mathbf{b}[i]$}. 

A convolution layer is another form of linear transformation that operates on images. The input of a \lname{CONV} layer is represented as multiple rectangular channels (2D images) of the same size: $\mathbf{X}\in \mathbb{R}^{h1 \times h2\times c}$, where $h1$ and $h2$ are the dimensions of the image and $c$ is the number of channels. The \lname{CONV} layer maps the input image into an output image $\mathbf{Y} \in \mathbb{R}^{h1^\prime \times h2^\prime \times f}$. A \lname{CONV} layer consists of a weight tensor $\mathbf{W} \in \mathbb{R}^{k\times k\times c \times f}$ and a bias vector $\mathbf{b}\in \mathbb{R}^{f}$. The $i$-th output channel in a \lname{CONV} layer is computed by sliding the kernel $\mathbf{W}[:,:,:,i] \in \mathbb{R}^{k\times k\times c}$ over the input, computing the dot product between the kernel and the windowed input, and adding the bias term $\mathbf{b}[i]$ to the result. 

\vspace{0.2em}
{\noindent \bf Non-linear Activations: }
The output of linear transformations (i.e., \lname{CONV} and \lname{FC}) is usually fed to an activation layer, which applies an element-wise non-linear transformation to the vector/tensor and generates an output with the same dimensionality. 
In this paper, we particularly utilize the Binary Activation (\lname{BA}) function for hidden layers. 
\lname{BA} maps the input operand to its sign value (i.e., $+1$ or $-1$).

\vspace{0.2em}
{\noindent \bf Batch Normalization: }
A batch normalization (\lname{BN}) layer is typically applied to the output of linear layers to normalize the results. If a \lname{BN} layer is applied to the output of a \lname{CONV} layer, it multiplies all of the $i$-th channel's elements by a scalar $\boldsymbol{\gamma}[i]$ and adds a bias term $\boldsymbol{\beta}[i]$ to the resulting channel. If BN is applied to the output of an \lname{FC} layer, it multiplies the $i$-th element of the vector by a scalar $\boldsymbol{\gamma}[i]$ and adds a bias term $\boldsymbol{\beta}[i]$ to the result.

\vspace{0.2em}
{\noindent \bf Pooling: }
Pooling layers operate on image channels outputted by the \lname{CONV} layers. A pooling layer slides a window on the image channels and aggregates the elements within the window into a single output element. Max-pooling and Average-pooling are two of the most common pooling operations in neural networks. Typically, pooling layers reduce the image size but do not affect the number of channels.

\vspace{-0.8em}
\subsection{Secret Sharing}\label{ssec:ss}
A secret can \rebut{be securely} shared among two or multiple parties using Secret Sharing (SS) schemes. 
An SS scheme guarantees that each share does not reveal any information about the secret.
The secret can be reconstructed using all (or subset) of shares.
In \sys{}, we use additive secret sharing in which a secret ${S}$ is shared among two parties by sampling a random number $\widehat{S}_1\in_R\mathbb{Z}_{2^b}$ (integers modulo $2^b$) as the first share and creating the second share as $\widehat{S}_2 = {S} - \widehat{S}_1\ \textit{mod}\ 2^b$ where $b$ is the number of bits to describe the secret. While none of the shares reveal any information about the secret ${S}$, they can be used to reconstruct the secret as ${S} = \widehat{S}_1 + \widehat{S}_2 \ \textit{mod}\ 2^b$. Suppose that two secrets ${S}^{(1)}$ and ${S}^{(2)}$ are shared among two parties where party-$1$ has $\widehat{S}^{(1)}_1$ and $\widehat{S}^{(2)}_1$ and party-$2$ has $\widehat{S}^{(1)}_2$ and $\widehat{S}^{(2)}_2$. Party-$i$ can create a share of the sum of two secrets as $\widehat{S}^{(1)}_i + \widehat{S}^{(2)}_i \ \textit{mod}\ 2^b$ without communicating to the other party. This can be generalized for arbitrary (more than two) number of secrets as well. 
\rebut{We utilize additive secret sharing in our Oblivious Conditional Addition (OCA) protocol (\sect{ssec:oca}).}

\vspace{-0.8em}
\subsection{Oblivious Transfer}\label{ssec:ot}
\rebut{One of} the most crucial building blocks of secure computation protocols, e.g., GC, is the Oblivious Transfer (OT) protocol~\cite{rabin2005exchange}. In OT, two parties are involved: a sender and a receiver. The sender holds $n$ different messages $m_j,\ j=1...n,$ with a specific bit-length and the receiver holds an index ($ind$) of a message that she wants to receive. At the end of the protocol, the receiver gets $m_{ind}$ with no additional knowledge about the other messages and the sender learns nothing about the selection index. In GC, 1-out-of-2 OT is used where $n=2$ in which case the selection index is only one bit. The initial realizations of OT required costly public key encryptions for each run of the protocol. However, the OT Extension~\cite{ishai2003extending,beaver1996correlated,asharov2013more} technique \rebut{enables performing} OT using more efficient symmetric-key encryption in conjunction with a {\it fixed} number of base OTs that need public-key encryption. 
\rebut{OT is used both in the OCA protocol as well as the Garbled Circuits protocol which we discuss next.}

\vspace{-0.8em}
\subsection{Garbled Circuits}\label{ssec:gc}
Yao's Garbled Circuits~\cite{yao1986generate}, or GC in short, is \rebut{one of} the generic two-party secure computation protocols. In GC, the result of an arbitrary function $f(.)$ on inputs from two parties can be computed without revealing each party's input to the other. 
Before executing the protocol, function $f(.)$ has to be described as a Boolean circuit with two-input gates.

GC has three main phases: garbling, transferring data, and evaluation. 
In the first phase, only one party, the Garbler, is involved. The Garbler starts by assigning two randomly generated $l$-bit binary strings to each wire in the circuit. These binary strings are called {\it labels} and they represent semantic values 0 and 1. Let us denote the label of wire $w$ corresponding to the semantic value $x$ as $L_x^w$. 
For each gate in the circuit, the Garbler creates a four-row garbled table as follows. Each label of the output wire is encrypted using the input labels according to the truth table of the gate. For example, consider an \texttt{AND} gate with input wires $a$ and $b$ and output wire $c$. The last row of the garbled table is the encryption of $L_1^c$ using labels $L_1^a$ and $L_1^b$. 

Once the garbling process is finished, the Garbler sends all of the garbled tables to the Evaluator. Moreover, he sends the correct labels that correspond to input wires that represent his inputs to the circuit. For example, if wire $w^*$ is the first input bit of the Garbler and his input is 0, he sends $L_0^*$. 
The Evaluator acquires the labels corresponding to her input through 1-out-of-2 OT where Garbler is the sender with two labels as his messages and the Evaluator's selection bit is her input for that wire. Having all of the garbled tables and labels of input wires, the Evaluator can start decrypting the garbled tables one by one until reaching the final output bits. She then learns the plaintext result at the end of the GC protocol based on the output labels and their relationships to the semantic values that are received from the Garbler. 

\section{The {\sys{}} Framework}\label{frmwk}
In this section, we explain how neural networks can be trained such that they incur \rebut{a minimal} cost during the oblivious inference. The most computationally intensive operation in a neural network is matrix multiplication. 
In GC, each multiplication has a {\it quadratic} computation and communication cost with respect to the input bit-length. This is the major source of inefficiency in prior work~\cite{deepsecure}. We overcome this limitation by changing the learning process such that the trained neural network's weights become binary. As a result, costly multiplication operations are replaced with \texttt{XNOR} gates which are essentially free in GC. We describe the training process in \sect{ssec:binarization}. In \sect{ssec:infer}, we explain the operations and their corresponding Boolean circuit designs that enable a very fast oblivious inference. In \sect{sec:imp}, we elaborate on \sys{} implementation. 

\vspace{-0.4em}
\subsection{Customized Network Binarization}\label{ssec:binarization}
Numerical optimization algorithms minimize a specific cost function associated with neural networks. It is well-known that neural network training is a non-convex optimization, meaning that there exist many locally-optimum parameter configurations that result in similar inference accuracies. Among these parameter settings, there exist solutions where both neural network parameters and activation units are restricted to take binary values (i.e., either $+1$ or $-1$); these solutions are known as Binary Neural Netowrks (BNNs)~\cite{courbariaux2016binarized}.

One major shortcoming of BNNs is their (often) low inference accuracy. In the machine learning community, several methods have been proposed to modify BNN functionality for accuracy enhancement~\cite{rastegari2016xnor,ghasemzadeh2018rebnet,lin2017towards}. These methods are devised for \textit{plaintext} execution of BNNs and are not efficient for oblivious inference with GC. We emphasize that, when modifying BNNs for accuracy enhancement, one should also take into account the implications in the corresponding GC circuit. With this in mind, we propose to modify the number of \rebut{channels and neurons} in \lname{CONV} and \lname{FC} layers, respectively. 
Increasing the number of channels/neurons leads to a higher accuracy but it also increases the complexity of the corresponding GC circuit. As a result, \sys{} provides a trade-off between the accuracy and the communication/runtime of the oblivious inference. This tradeoff enables cloud servers to customize the complexity of the GC protocol to optimally match the computation and communication requirements of the clients.
\rebut{To customize the BNN, \sys{} configures the per-layer number of neurons in two steps:}


\begin{itemize}[leftmargin=*]
\item Linear Scaling: \rebut{Prior to training, we scale the number of channels/neurons in all BNN layers with the same factor ($s$), e.g., $s=2$. Then, we train the scaled BNN architecture.}
\item Network Trimming: Once the (uniformly) scaled network is trained, a post-processing algorithm removes redundant channels/neurons from each hidden layer to reduce the GC cost while maintaining the inference accuracy.
\end{itemize}

\begin{figure}
    \centering
    \includegraphics[width=\columnwidth]{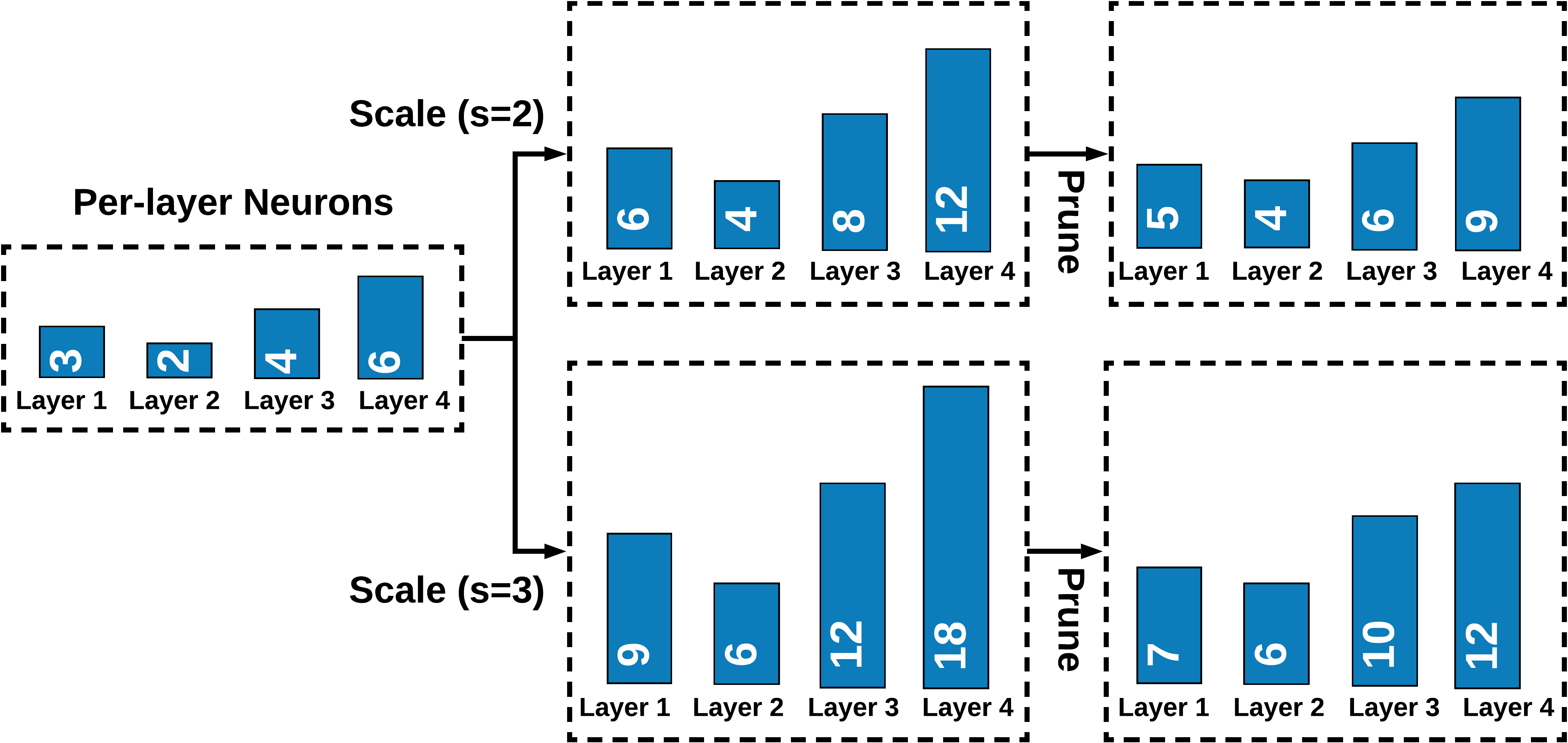}
    \caption{Illustration of BNN customization. The bars represent the number of neurons in each hidden layer.}
    \vspace{-0.5cm}
    \label{fig:cnn_prune}
\end{figure}

Figure~\ref{fig:cnn_prune} illustrates the BNN customization method for an example baseline network with four hidden layers.
\rebut{Network trimming (pruning) consists of two steps, namely, Feature Ranking and Iterative Pruning which we describe next.}

\vspace{0.2em}
\noindent{\bf Feature Ranking: }
In order to perform network trimming, one needs to sort the channels/neurons of each layer based on their contribution to the inference accuracy. In conventional neural networks, simple ranking methods sort features based on absolute value of the neurons/channels~\cite{han2015learning}. In BNNs, however, the weights/features are either $+1$ or $-1$ and the absolute value is not informative. To overcome this issue, we utilize first order Taylor approximation of neural networks and sort the features based on the magnitude of the gradient values~\cite{molchanov2016pruning}. Intuitively, the gradient with respect to a certain feature determines its importance; a high (absolute) gradient indicates that removing the neuron has a destructive effect on the inference accuracy. Inspired by this notion, we develop a feature ranking method described in Algorithm~\ref{alg:ranking_conv}. 

\vspace{0.2em}
\noindent{\bf Iterative Pruning:} We devise a step-by-step algorithm for model pruning which is summarized in Algorithm~\ref{alg:iterative_pruning}. At each step, the algorithm selects one of the BNN layers $l^*$ and removes the first $p^*$ features with the lowest importance (line~\ref{line:prune_step}). The selected layer $l^*$ and the number of pruned neurons $p^*$ maximize the following reward (line~\ref{line:optimal}):
\begin{equation}
   \mli{reward}(l,p)=\frac{c_{curr}-c_{next}}{e^{a_{curr}-a_{next}}},
\end{equation}
where $c_{curr}$ and $c_{next}$ are the GC complexity of the BNN before and after pruning, whereas, $a_{curr}$ and $a_{next}$ denote the corresponding validation accuracies. The numerator of this reward encourages higher reduction in the GC cost while the denominator penalizes accuracy loss. 
Once the layer is pruned, the BNN is fine-tuned to recover the accuracy (line~\ref{line:finetune}). 
The pruning \rebut{process stops} once the accuracy drops below a pre-defined threshold.

\begin{algorithm} [h]
        \caption{\sys{} Channel Sorting for \lname{CONV} Layers}\label{alg:ranking_conv}
\begin{flushleft}
        \textbf{Inputs:} Trained BNN with loss function $\mathcal{L}$, \lname{CONV} layer $l$ with output shape of $h1\times h2\times f$, subsampled validation data and labels $\{(\mathbf{X_1},z_1), \dots, (\mathbf{X_k}, z_k)\}$
        
        \textbf{Output:} Indices of the sorted channels: $\{i_0, \dots, i_f\}$
\end{flushleft}
\begin{algorithmic}[1]
\State $\mathbf{G}\gets \mli{zeros}(k\times h1\times h2\times f)$ \Comment{define gradient tensor}
\For {$i = 1, \dots, k$}
    \State $\mathcal{L}=\mathcal{L}(\mathbf{X_i},z_i)$ \Comment{evaluate loss function}
    \State $\nabla_Y=\frac{\partial \mathcal{L}}{\partial Y^l}$ \Comment{compute gradient w.r.t. layer output}
    \State $\mathbf{G}[i,:,:,:]\gets\nabla_Y$\Comment{store gradient}
\EndFor
\State $\mathbf{G_{abs}}\gets|\mathbf{G}|$ \Comment{take elementwise absolute values}
\State $\mathbf{g_s}\gets \mli{zeros}(f)$ \Comment{define sum of absolute values}
\For {$i = 1, \dots, f$}
    \State $\mathbf{g_s}[i]\gets sum(\mathbf{G_{abs}}[:,:,:,i])$
\EndFor

\State $\{i_0, \dots, i_f\} \gets \mli{sort}(\mathbf{g_s})$
\State \textbf{return} $\{i_0, \dots, i_f\}$
\end{algorithmic}
\end{algorithm}

\begin{algorithm*}[ht]
\caption{\sys{} Iterative BNN Pruning}\label{alg:iterative_pruning}
\begin{flushleft}
        \textbf{Inputs:} Trained BNN with $n$ overall \lname{CONV} and \lname{FC} layers, minimum accuracy threshold $\theta$, number of pruning trials per layer $t$, subsampled validation data and labels $\mli{data_{V}}$, training data and labels $\mli{data_{T}}$
        
        \textbf{Output:} BNN with pruned layers
\end{flushleft}
\begin{algorithmic}[1]
\State $\mathbf{p}\gets \mli{zeros}(n-1)$ \Comment{current number of pruned neurons/channels per layer}
\State $a_{curr}\gets Accuracy(\mli{BNN},\mli{data_{V}}|\mathbf{p})$ \Comment{current BNN validation accuracy}
\State $c_{curr}\gets Cost(\mli{BNN}|\mathbf{p})$ \Comment{current GC cost}

\While{$a_{curr}>\theta$}\Comment{repeat until accuracy drops below $\theta$}
    \For{$l=1, \dots, n-1$}\Comment{search over all layers}
        \State $\mli{inds}\gets Rank(\mli{BNN}, l, \mli{data_{V}})$ \Comment{rank features via Algorithm~\ref{alg:ranking_conv}}
        \State $f \gets$ Number of neurons/channels\Comment{number of output neurons/channels}
        \For{$p=\mathbf{p}[l], \mathbf{p}[l]+\frac{f}{t}, \dots, f$}\Comment{search over possible pruning rates}
            \State $\mli{BNN_{next}}\gets Prune(\mli{BNN},l,p,\mli{inds})$ \Comment{prune $p$ features with lowest ranks from the $l$-th layer}\label{line:prune}
            \State $a_{next}\gets Accuracy(\mli{BNN_{next},\mli{data_V}}|\mathbf{p[1]}, \dots,\mathbf{p[l]}=p,\dots,\mathbf{p[n-1]})$ \Comment{validation accuracy if pruned}
            \State $c_{next}\gets Cost(\mli{BNN_{next}}|\mathbf{p[1]}, \dots,\mathbf{p[l]}=p,\dots,\mathbf{p[n-1]})$ \Comment{GC cost if pruned}
            \State $\mli{reward}(l,p)=\frac{c_{curr}-c_{next}}{e^{(a_{curr}-a_{next})}}$\Comment{compute reward given that $p$ features are pruned from layer $l$}
        \EndFor
    \EndFor
    \State $\{l^*,p^*\}\gets \argmax_{l,p}\mli{reward(l,p)}$\Comment{select layer $l^*$ and pruning rate $p^*$ that maximize the reward}\label{line:optimal}
    \State $\mathbf{p[l^*]}\gets p^*$\Comment{update the number of pruned features in vector $\mathbf{p}$}
    \State $\mli{BNN}\gets Prune(\mli{BNN},l^*,p^*,\mli{inds})$ \Comment{prune $p^*$ features with lowest ranks from the $l^*$-th layer}\label{line:prune_step}
    \State $\mli{BNN}\gets$ $\mli{\textit{Fine-tune}}(\mli{BNN}, \mli{data_{T}})$ \Comment{fine-tune the pruned model using training data to recover accuracy}\label{line:finetune}
    \State $a_{curr}\gets Accuracy(\mli{BNN},\mli{data_{V}}|\mathbf{p})$ \Comment{update current BNN validation accuracy}
    \State $c_{curr}\gets Cost(\mli{BNN}|\mathbf{p})$ \Comment{update current GC cost}
\EndWhile
\State \textbf{return} $\mli{BNN}$
\end{algorithmic}
\end{algorithm*}

\vspace{-0.7em}
\subsection{Oblivious Inference}\label{ssec:infer}
BNNs are trained such that the weights and activations are binarized, i.e., they can only have two possible values: $+1$ or $-1$. This property allows BNN layers to be rendered using a simplified arithmetic. In this section, we describe the functionality of different layer types in BNNs and their Boolean circuit translations. Below, we explain each layer type. 

\vspace{0.2em}
{\noindent \bf Binary Linear Layer: }
Most of the computational complexity of neural networks is due to the linear operations in \lname{CONV} and \lname{FC} layers. As we discuss in Section~\ref{ssec:dnn}, linear operations are realized using vector dot product (\lname{VDP}). In BNNs, VDP operations can be implemented using simplified circuits. We categorize the \lname{VDP} operations of this work into two classes: (i)~Integer-\lname{VDP} where only one of the vectors is binarized and the other has integer elements and (ii)~Binary-\lname{VDP} where both vectors have binary ($\pm1$) values.

\noindent{\it Integer-\lname{VDP}}: For the first layer of the neural network, the server has no control over the input data which is not necessarily binarized. The server can only train binary weights and use them for oblivious inference. Consider an input vector $\mathbf{x}\in \mathbb{R}^n$ with integer (possibly fixed-point) elements and a weight vector $\mathbf{w}\in\{-1,1\}^n$ with binary values. Since the elements of the binary vector can only take $+1$ or $-1$, the Integer-\lname{VDP} can be rendered using additions and subtractions. In particular, the binary weights can be used in a selection circuit that decides whether the pertinent integer input should be added to or subtracted from the \lname{VDP} result.     

\begin{figure}[h]
    \centering
    \includegraphics[width=0.98\columnwidth]{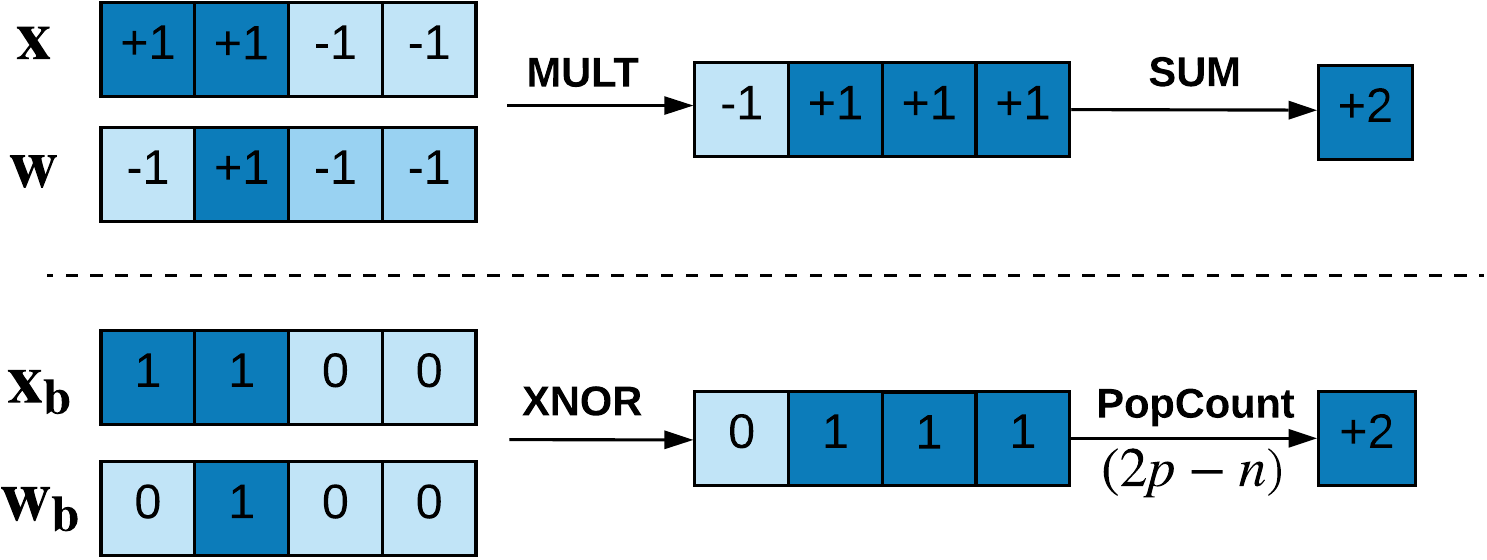}
    \caption{Equivalence of Binary-\lname{VDP} and $\mathbf{XnorPopcount}$. 
    }
    \vspace{-0.2em}
    \label{fig:xnorpopcount}
\end{figure}

\vspace{0.3em}
\noindent{\it Binary-\lname{VDP}}: Consider a dot product between two binary vectors $\mathbf{x}\in\{-1,+1\}^n$ and $\mathbf{w}\in\{-1,+1\}^n$. If we encode each element with one bit (i.e., $-1\rightarrow 0$ and $+1\rightarrow 1$), we obtain binary vectors $\mathbf{x}_b\in\{0,1\}^n$ and $\mathbf{w}_b\in\{0,1\}^n$. It has been shown that the dot product of $\mathbf{x}$ and $\mathbf{w}$ can be efficiently computed using an $XnorPopcount$ operation~\cite{courbariaux2016binarized}. \fig{fig:xnorpopcount} depicts the equivalence of $\lname{VDP}(\mathbf{x},\mathbf{w})$ and $XnorPopcount(\mathbf{x}_b,\mathbf{w}_{b})$ for a \lname{VDP} between $4$-dimensional vectors. First, element-wise \texttt{XNOR} operations are performed between the two binary encodings. Next, the number of set bits $p$ is counted, and the output is computed as $2p-n$.

\vspace{0.4em}
{\noindent \bf Binary Activation Function:} A Binary Activation (\lname{BA}) function takes input $x$ and maps it to $y=Sign(x)$ where $Sign(\cdot)$ outputs either $+1$ or $-1$ based on the sign of its input. This functionality can simply be implemented by extracting the most significant bit of $x$.

\vspace{0.4em}
{\noindent \bf Binary Batch Normalization:} in BNNs, it is often useful to normalize \rebut{feature} $x$ using a Batch Normalization (\lname{BN}) layer before applying the binary activation function. More specifically, a \lname{BN} layer followed by a \lname{BA} is equivalent to: $$y=Sign(\gamma\cdot x +\beta)=Sign(x+\frac{\beta}{\gamma}),$$
since $\gamma$ is a positive value. The combination of the two layers (\lname{BN+BA}) is realized by a comparison between $x$ and $-\frac{\beta}{\gamma}$.

\begin{figure}[h]
    \centering
    \includegraphics[width=\columnwidth]{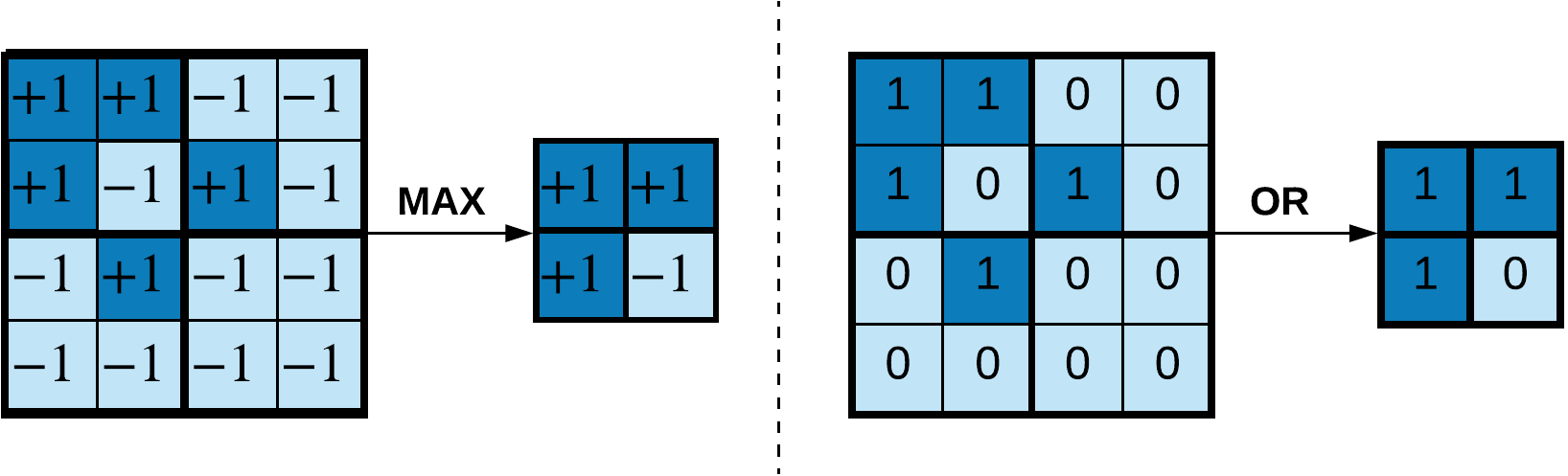}
    \caption{The equivalence between Max-Pooling and Boolean-\texttt{OR} operations in BNNs. }
    \label{fig:max_pool_OR}
    \vspace{-0.5cm}
\end{figure}

\vspace{0.3em}
{\noindent \bf Binary Max-Pooling:} Assuming the inputs to the max-pooling layers are binarized, taking the maximum in a window is equivalent to performing logical \texttt{OR} over the binary encodings as depicted in \fig{fig:max_pool_OR}. Note that average-pooling layers are usually not used in BNNs since the average of multiple binary elements is no longer a binary value.

\begin{figure}
    \centering
    \includegraphics[width=0.97\columnwidth]{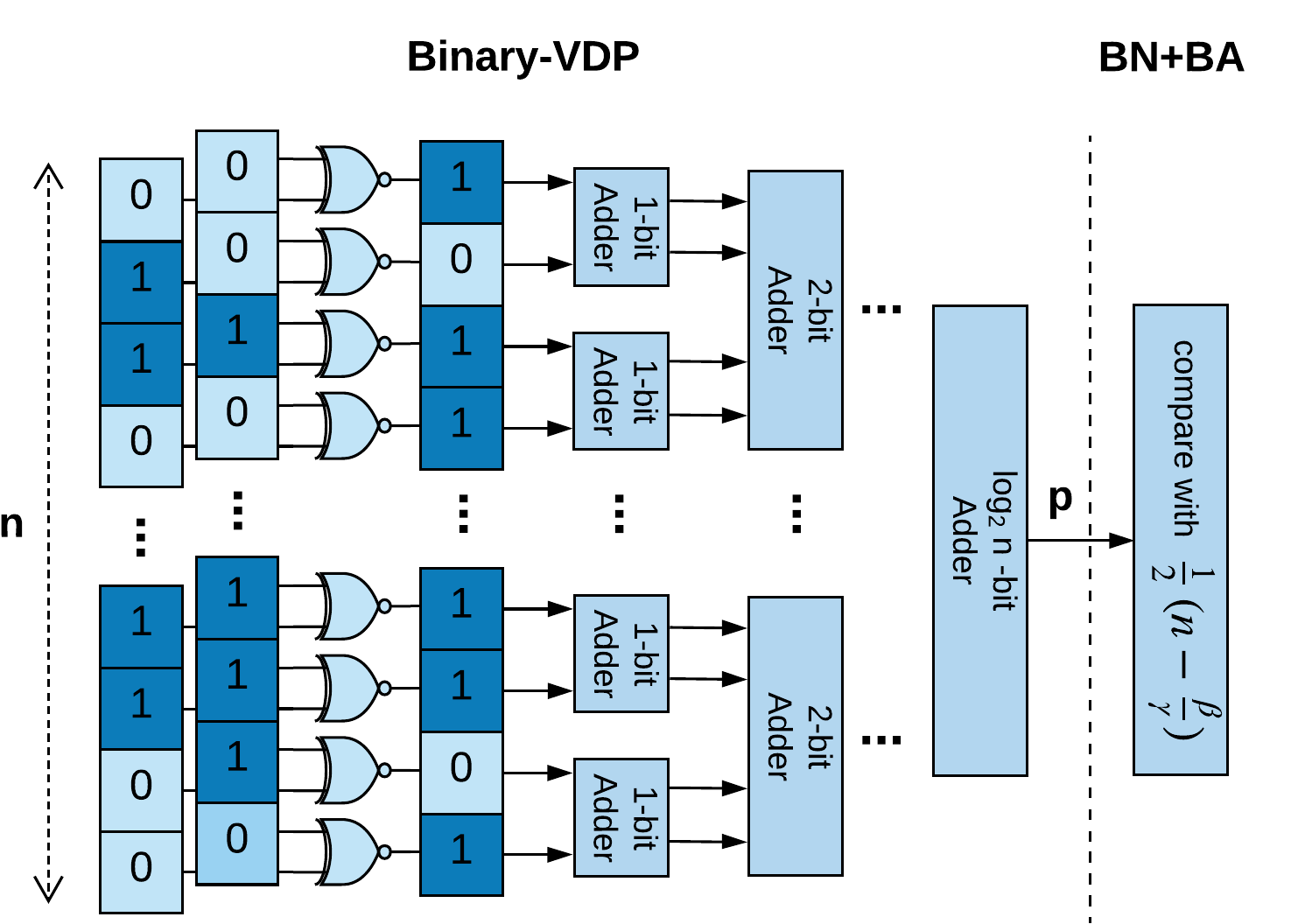}
    \caption{Circuit for binary-\lname{VDP} followed by comparison for batch normalization (\lname{BN}) and binary activation (\lname{BA}).
    }
    \label{fig:circuit_B_VDP}
\end{figure}

\begin{figure}[ht]
    \centering
    \includegraphics[width=\columnwidth]{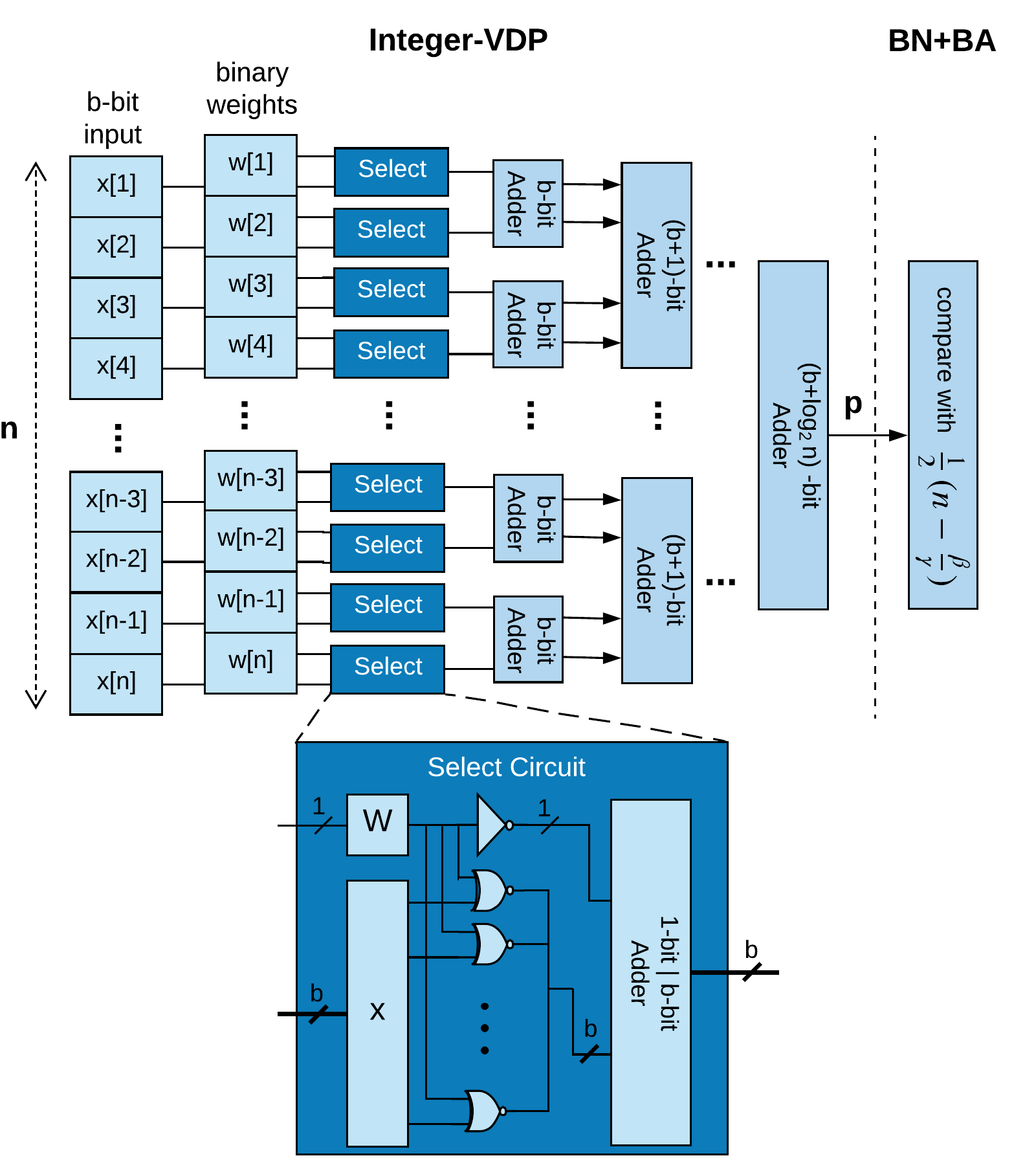}
    \caption{Circuit for Integer-\lname{VDP} followed by comparison for batch normalization (\lname{BN}) and binary activation (\lname{BN}).
    }
    \label{fig:circuit_I_VDP}
\end{figure}

\vspace{-0.2cm}
\fig{fig:circuit_B_VDP} demonstrates the Boolean circuit for Binary-\lname{VDP} followed by \lname{BN} and \lname{BA}. 
The number of non-\texttt{XOR} gates for binary-\lname{VDP} is equal to the number of gates required to render the tree-adder structure in \fig{fig:circuit_B_VDP}. 
Similarly, \fig{fig:circuit_I_VDP} shows the Integer-\lname{VDP} counterpart.
In the first level of the tree-adder of Integer-\lname{VDP} (\fig{fig:circuit_I_VDP}), the binary weights determine whether the integer input should be added to or subtracted from the final result within the ``Select'' \rebut{circuit}. The next levels of the tree-adder compute the result of the integer-\lname{VDP} using ``Adder'' blocks.
The combination of \lname{BN} and \lname{BA} is implemented using a single {\it comparator}. 
Compared to Binary-\lname{VDP}, Integer-\lname{VDP} has a high garbling cost which is linear with respect to the number of bits. To mitigate this problem, we propose an alternative solution based on Oblivious Transfer (OT) in~\sect{ssec:oca}.

\begin{figure}[ht]
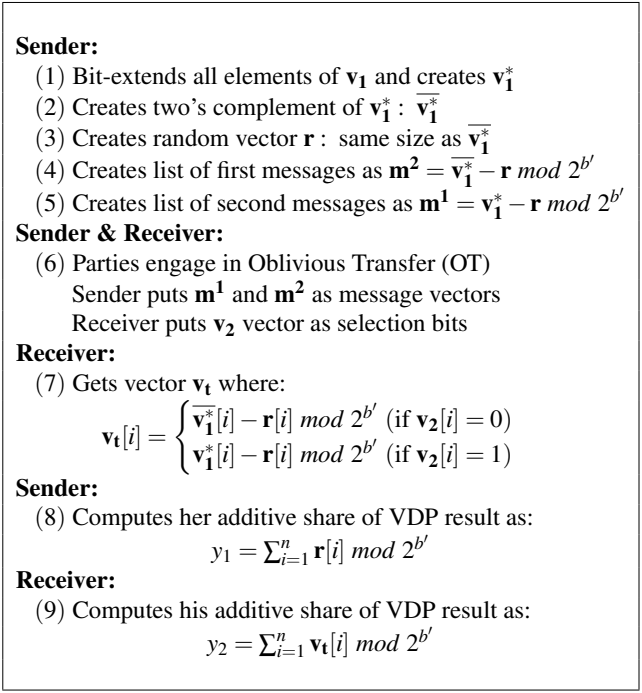

\centering
\[\resizebox{\columnwidth}{!}{$\displaystyle
\begin{array}{|l|}\hline\\
\textbf{Sender:}\\
~~~(1)\ \text{Bit-extends all elements of}\ \mathbf{v_1}\ \text{and creates}\ \mathbf{v^*_1}\\
~~~(2)\ \text{Creates two's complement of}\ \mathbf{v^*_1}:\ \overline{\mathbf{v^*_1}}\\
~~~(3)\ \text{Creates random vector}\ \mathbf{r}:\ \text{same size as}\ \overline{\mathbf{v^*_1}}\\
~~~(4)\ \text{Creates list of first messages as}\ \mathbf{m^2 =}\ \overline{\mathbf{v^*_1}}-\mathbf{r}\ mod\ 2^{b'}\\
~~~(5)\ \text{Creates list of second messages as}\ \mathbf{m^1 = v^*_1-r}\ mod\ 2^{b'}\\
\textbf{Sender \& Receiver:}\\
~~~(6)\ \text{Parties engage in Oblivious Transfer (OT)}\\ 
~~~~~~~~~\text{Sender puts}\ \mathbf{m^1}\ \text{and}\ \mathbf{m^2} \text{ as message vectors}\\
~~~~~~~~~\text{Receiver puts}\ \mathbf{v_2}\ \text{vector as selection bits}\\
\textbf{Receiver:}\\
~~~(7)\ \text{Gets vector}\ \mathbf{v_t}\ \text{where:}\\
\mc{\mathbf{v_t}[i]=\begin{cases}
      \overline{\mathbf{v^*_1}}[i]-\mathbf{r}[i]\ mod\ 2^{b'}\ (\text{if}\ \mathbf{v_2}[i]=0)\\
      \mathbf{v^*_1}[i]-\mathbf{r}[i]\ mod\ 2^{b'}\ (\text{if}\ \mathbf{v_2}[i]=1)\\
   \end{cases}}\\
\textbf{Sender:}\\
~~~(8)\ \text{Computes her additive share of VDP result as:}\\ \mc{y_1=\sum_{i=1}^{n} \mathbf{r}[i] \ mod\ 2^{b'}}\\
\textbf{Receiver:}\\
~~~(9)\ \text{Computes his additive share of VDP result as:}\\ \mc{y_2=\sum_{i=1}^{n} \mathbf{v_t}[i] \ mod\ 2^{b'}}\\\\\hline
\end{array}
$}\]
\vspace{-0.5cm}
\caption{Oblivious Conditional Addition (OCA) protocol.}
\label{fig:oca}
\vspace{-0.2em}
\end{figure}

\subsection{Oblivious Conditional Addition Protocol}\label{ssec:oca}
In \sys{}, all of the activation values as well as neural network weights are binary. \rebut{However, the input to the neural network is provided by the user and is not necessarily binary}.
The first layer of a typical neural network comprises either an \lname{FC} or a \lname{CONV} layer, both of which are evaluated using oblivious Integer-\lname{VDP}. On the one side, the user provides her input as non-binary (integer) values. On the other side, the network parameters are binary values representing $-1$ and $1$. \rebut{We now demonstrate} how Integer-\lname{VDP} can be described as an OT problem. Let us denote the user's input as a vector $\mathbf{v_1}$ of $n$ ($b$-bit) integers. The server holds a vector of $n$ binary values denoted by $\mathbf{v_2}$. The result of Integer-\lname{VDP} is a number ``$y$'' that can be described with
$$b'=\ceil*{\text{log}_2(n\cdot (2^b-1))}$$ 
bits. \fig{fig:oca} summarizes the steps in the OCA protocol. The first step is to {\it bit-extend} $\mathbf{v_1}$ from $b$-bit to $b'$-bit. In other words, if $\mathbf{v_1}$ is a vector of {\it signed} integer/fixed-point numbers, the most significant bit should be repeated \rebut{($b'- b$)-many times}, otherwise, it has to be zero-padded for most significant bits. We denote the bit-extended vector by $\mathbf{v^*_1}$. The second step is to create the two's complement vector of $\mathbf{v^*_1}$, called $\overline{\mathbf{v^*_1}}$. The client also creates a vector of $n$ ($b'$-bit) randomly generated numbers, denoted as $\mathbf{r}$. She computes element-wise vector subtractions $\mathbf{v^*_1-r}\ mod\ 2^{b'}$ and $\overline{\mathbf{v^*_1}}-\mathbf{r}\ mod\ 2^{b'}$. These two vectors \rebut{are $n$-many pair} of messages that \rebut{will be used as} input to $n$-many 1-out-of-two OTs.\rebut{ More precisely, $\overline{\mathbf{v^*_1}}-\mathbf{r}\ mod\ 2^{b'}$ is a list of first messages and $\mathbf{v^*_1-r}\ mod\ 2^{b'}$ is a list of second messages.} 
The server's list of selection bits is $\mathbf{v_2}$. After $n$-many OTs are finished, the server has a list of $n$ transferred numbers called $\mathbf{v_t}$ where 

\[\mathbf{v_t}[i] = \left\{
                \begin{array}{ll}
                  \overline{\mathbf{v^*_1}}[i]-\mathbf{r}[i]\ mod\ 2^{b'}\ \ \ \ if\ \ \mathbf{v_2}[i]=0\\
                  \mathbf{v^*_1}[i]-\mathbf{r}[i]\ mod\ 2^{b'}\ \ \ \ if\ \ \mathbf{v_2}[i]=1\\
                \end{array}
              \right.\ i=1,\ ...\ ,\ n.
\]

Finally, the client computes $y_1=\sum_{i=1}^{n} \mathbf{r}[i] \ mod\ 2^{b'}$ and the server computes $y_2=\sum_{i=1}^{n} \mathbf{v_t}[i] \ mod\ 2^{b'}$. By OT's definition, the receiver (server) gets only one of the two messages from the sender. That is, based on each selection bit (a binary weight), the receiver gets an additive share of either the sender's number or its two's complement. Upon adding all of the received numbers, the receiver computes an additive share of the Integer-\lname{VDP} result. Now, even though the sender does not know which messages were selected by the receiver, she can add all of the randomly generated numbers $\mathbf{r}[i]$s which is equal to the other additive share of the Integer-\lname{VDP} result. 
Since all numbers are described in the two's complement format, subtractions are equivalent to the addition of the two's complement values, which are created by the sender at the beginning of OCA. Moreover, it is possible that as we accumulate the values, the bit-length of the final Integer-\lname{VDP} result grows accordingly. This is \rebut{supported due to the} bit-extension process at the beginning of the protocol. In other words, all additions are performed in a larger ring such that the result does not overflow. 

Note that all numbers belong to the ring $\mathbb{Z}_{2^{b'}}$ and by definition, a ring is closed under addition, therefore, $y_1$ and $y_2$ are true additive shares of $y=y_1+y_2\ mod\ 2^{b'}$.
We described the OCA protocol for one Integer-\lname{VDP} computation. As we outlined in \sect{ssec:infer}, all linear operations in the first layer of the DL model \rebut{(either \lname{FC} or \lname{CONV})} can be formulated as a series of Integer-\lname{VDP}s. 


In traditional OT, public-key encryption is needed for each OT invocation which can be computationally expensive.
Thanks to the Oblivious Transfer Extension technique~\cite{ishai2003extending,beaver1996correlated,asharov2013more}, one can perform many OTs using symmetric-key encryption and only a fixed number of public-key operations.


\vspace{0.2em}
\noindent {\bf Required Modification to the Next Layer.} So far, we have shown how to perform Integer-\lname{VDP} using OT. However, we need to add an ``addition'' layer to reconstruct the true value of $y$ from its additive shares before further processing it. The overhead of this layer, as well as OT computations, are discussed next. Note that OCA is used only for the first layer and it does not change the overall constant round complexity of \sys{} since it is performed only once regardless of the number of layers in the DL model.

\vspace{0.2em}
\noindent {\bf Comparison to Integer-VDP in GC.}
\tab{tab:oca} shows the computation and communication costs for two approaches: (i) computing the first layer in GC and (ii) utilizing OCA. 
OCA removes the GC cost of the first layer in \sys{}. However, it adds the overhead of a set of OTs and the GC costs associated with the new \lname{ADD} layer. 

\vspace{-0.2em}
\begin{table}[h]
\centering
\caption{Computation and communication cost of OCA.}
\vspace{-0.3em}
\label{tab:oca}
\resizebox{0.99\columnwidth}{!}{
\begin{tabular}{lccc}\toprule
\multirow{2}{*}{\begin{tabular}[c]{@{}c@{}}\textbf{Costs}\\ \{Sender,\ Receiver\}\end{tabular}}              & \multirow{2}{*}{\textbf{GC}} & \multicolumn{2}{c}{\textbf{OCA}}          \\
             &                     & OT              & \texttt{ADD} Layer             \\\midrule
Comp. (AES ops)   & $(n+1) \cdot b\ \cdot$ \{2,\ 4\}  & $n\ \cdot$ \{1,\ 2\} & $b'\cdot$ \{2,\ 4\} \\
Comm. (bit) & $(n+1)\cdot b\cdot 2\cdot 128$        & $n\cdot b$             & $b'\cdot 2\cdot 128$  \\\bottomrule    
\end{tabular}
}
\end{table}

\vspace{-0.7em}
\subsection{Security of \sys{}}\label{ssec:sec}
We consider the Honest-but-Curious (HbC) adversary model consistent with all of the state-of-the-art solutions for oblivious inference~\cite{deepsecure,secureml,minionn,chameleon,ezpc,gazelle}. In this model, neither of the \rebut{involved} parties is trusted but they are assumed to follow the protocol. Both server and client cannot infer any information about the other party's input from the entire protocol transcript. 
\sys{} relies solely on the GC and OT protocols, both of which are proven to be secure in the HbC adversary model in~\cite{lindell2009proof} and~\cite{rabin2005exchange}, respectively. 
Utilizing binary neural networks does not affect GC and OT protocols in any way. More precisely, we have changed the function $f(.)$ that is evaluated in GC \rebut{such that} it is more efficient for the GC protocol: drastically reducing the number of \texttt{AND} gates and using \texttt{XOR} gates instead. 
Our novel Oblivious Conditional Addition (OCA) protocol (\sect{ssec:oca}) is also based on the OT protocol. The sender creates a list of message pairs and puts them as input to the OT protocol. 
Each message is an additive share of the sender's private data from which the secret data cannot be reconstructed. 
The receiver puts a list of selection bits as input to the OT. By OT's definition, the receiver learns nothing about the unselected messages and the sender does not learn the selection bits. 

During the past few years, several attacks have been proposed that extract some information about the DL model by querying the server many times~\cite{tramer2016stealing,fredrikson2015model,shokri2017membership}. It has been shown that some of these attacks can be effective in the black-box setting where the client only receives the prediction results and does not have access to the model. Therefore, considering the definition of an oblivious inference, these type of attacks are out of the scope of oblivious inference frameworks. However, in~\app{sec:attacks}, we show how these attacks can be thwarted by adding a simple layer at the end of the neural network which adds a negligible overhead. 

\vspace{0.2em}
\noindent{\bf Security Against Malicious Adversaries.}
The HbC adversary model is the standard security model in the literature. However, there are more powerful security models such as security against covert and malicious adversaries. In the malicious security model, the adversary (either the client or server) can deviate from the protocol at any time with the goal of learning more about the input from the other party. 
One of the main distinctions between \sys{} and the state-of-the-art solutions is that \sys{} can be automatically \rebut{adapted} to the malicious security using cut-and-choose techniques~\cite{lindell2012secure,huang2013efficient,lindell2016fast}. These methods take a GC protocol in HbC and readily extend it to the malicious security model. This modification increases the overhead but enables a higher security level.
To the best of our knowledge, there is no practical solution to extend the customized mixed-protocol frameworks~\cite{minionn,chameleon,ezpc,gazelle} to the malicious security model.
Our GC-based solution is more efficient compared to the mixed-protocol solutions and can be upgraded to the malicious security at the same time. 

\vspace{-0.4em}
\section{The \sys{} Implementation}\label{sec:imp}
\vspace{-0.2em}
In this section, we elaborate on the garbling/evaluation implementation of \sys{}. 
All of the optimizations and techniques proposed in this section do not change the security \rebut{or} correctness in anyway and only enable the framework's scalability for large network architectures. 

We design a new GC framework with the following design principles in mind: 
(i) {\it Efficiency:} \sys{} is designed to have a minimal data movement and low cache-miss rate. 
(ii) {\it Scalability: } oblivious inference inevitably requires significantly higher memory usage compared to plaintext evaluation of neural networks. High memory usage is one critical shortcoming of state-of-the-art secure computation frameworks. As we show in our experimental results, \sys{} is designed to scale for very deep neural networks that have higher accuracy compared to networks considered in prior art. 
(iii) {\it Modularity:} our framework enables users to create Boolean description of different layers separately. This allows the hardware synthesis tool to generate more optimized circuits as we discuss in~\sect{ssec:syn}. 
(iv) {\it Ease-to-use:} \sys{} provides a very simple API that requires few lines of neural network description. Moreover, we have created a compiler that takes a Keras description and automatically creates the network description for \sys{} API. 

\sys{} is written in C++ and supports all major GC optimizations proposed previously. Since the introduction of GC, many optimizations have been proposed to reduce the computation and communication complexity of this protocol.  Bellare et al.~\cite{bellare2013efficient} have provided a way to perform garbling using efficient fixed-key AES encryption. Our implementation benefits from this optimization by using Intel AES-NI instructions. 
Row-reduction technique~\cite{naor1999privacy} reduces the number of garbled tables from four to three. 
Half-Gates technique~\cite{zahur2015two} further reduces the number of rows in the garbled tables from three to two. 
One of the most influential optimizations for the GC protocol is the {\it free-XOR} technique~\cite{freexor} which makes \texttt{XOR}, \texttt{XNOR}, and \texttt{NOT} almost free of cost. 
Our implementation for Oblivious Transfer (OT) is based on libOTe~\cite{libOTe}.

\begin{figure}[h]
    \centering
    \includegraphics[width=0.95\columnwidth]{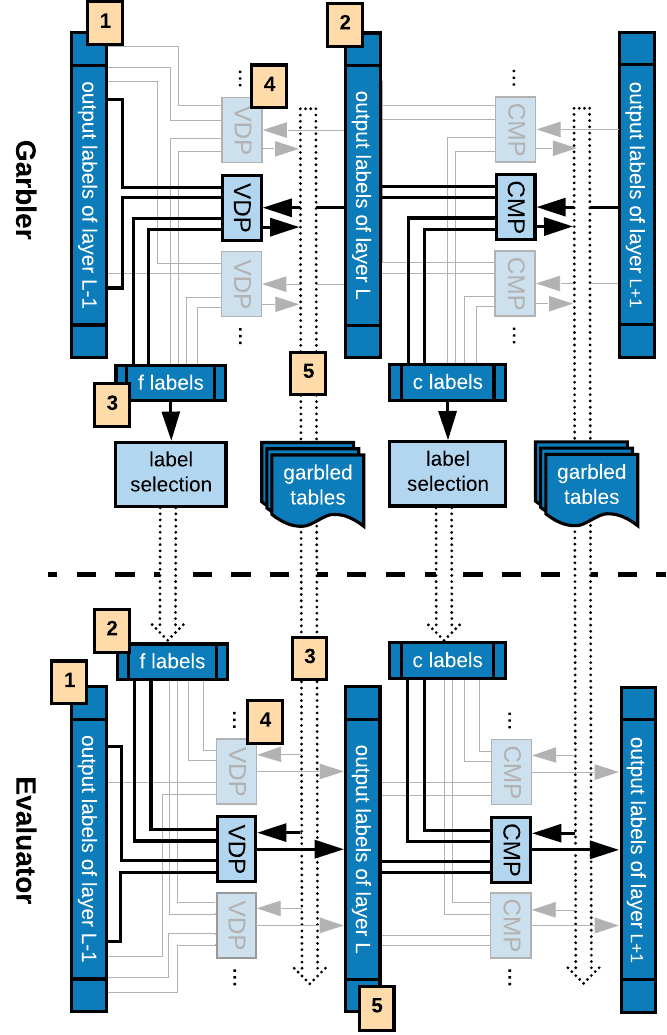}
    \caption{\sys{} modular and pipelined garbling engine.}
    \label{fig:label}
\end{figure}

\vspace{-1em}
\subsection{Modular Circuit Synthesis and Garbling}\label{ssec:syn}
In \sys{}, each layer is described as multiple invocations of a {\it base} circuit. For instance, linear layers (\lname{CONV} and \lname{FC}) are described by a \lname{VDP} circuit. \lname{MaxPool} is described by an \texttt{OR} circuit where the number of inputs is the window size of the \lname{MaxPool} layer. \lname{BA}/\lname{BN} layers are described using a comparison (\texttt{CMP}) circuit. The memory footprint is significantly reduced in this approach: we only create and store the base circuits. As a result, the connection between two invocations of two different base circuits is handled at the software level. 

We create the Boolean circuits using TinyGarble~\cite{songhori2015tinygarble} hardware synthesis approach. TinyGarble's technology libraries are optimized for GC and produce circuits that have low number of non-\texttt{XOR} gates. 
Note that the Boolean circuit description of the contemporary neural networks comprises between millions to billions of Boolean gates, whereas, synthesis tools cannot support circuits of this size. 
However, due to \sys{} modular design, \rebut{one can synthesize} each base circuit separately. 
Thus, the bottleneck transfers from the synthesis tool's maximum number of gates to the system's memory. 
As such, \sys{} effectively scales for any neural network complexity regardless of the limitations of the synthesis tool as long as enough memory (i.e., RAM) is available.
Later in this section, we discuss how to increase the scalability by dynamically managing the allocated memory. 

\vspace{0.3em}
\noindent {\bf Pipelined GC Engine.} 
In \sys{}, computation and communication are pipelined. 
For instance, consider a \lname{CONV} layer followed by an activation layer. We garble/evaluate these layers by multiple invocations of the \lname{VDP} and \texttt{CMP} circuits (one invocation per output neuron) as illustrated in \fig{fig:label}. Upon finishing the garbling process of layer $L-1$, the Garbler starts garbling the $L^{th}$ layer and creates the random labels for output wires of layer $L$. He also needs to create the random labels associated with his input (i.e., the weight parameters) to layer $L$. Given a set of input and output labels, Garbler generates the garbled tables, and sends them to the Evaluator as soon as one is ready. He also sends one of the two input labels for his input bits. At the same time, the Evaluator has computed the output labels of the $(L-1)^{th}$ layer. She receives the garbled tables as well as the Garbler's selected input labels and \rebut{decrypts} the tables and stores the output labels of layer $L$.  

\vspace{0.3em}
\noindent {\bf Dynamic Memory Management.} We design the framework such that the allocated memory for the labels is released as soon as it is no longer needed, reducing the memory usage significantly. 
For example, without our dynamic memory management, the Garbler had to allocate $10.41$GB for the labels and garbled tables for the entire garbling of BC1 network \rebut{(see \sect{sec:eval} for network description)}. In contrast, in our framework, the size of memory allocation never exceeds 2GB and is less than 0.5GB for most of the layers.  


\subsection{Application Programming Interface (API)}\label{ssec:api}
\begin{figure}[h]
    \centering
    \includegraphics[width=0.95\columnwidth]{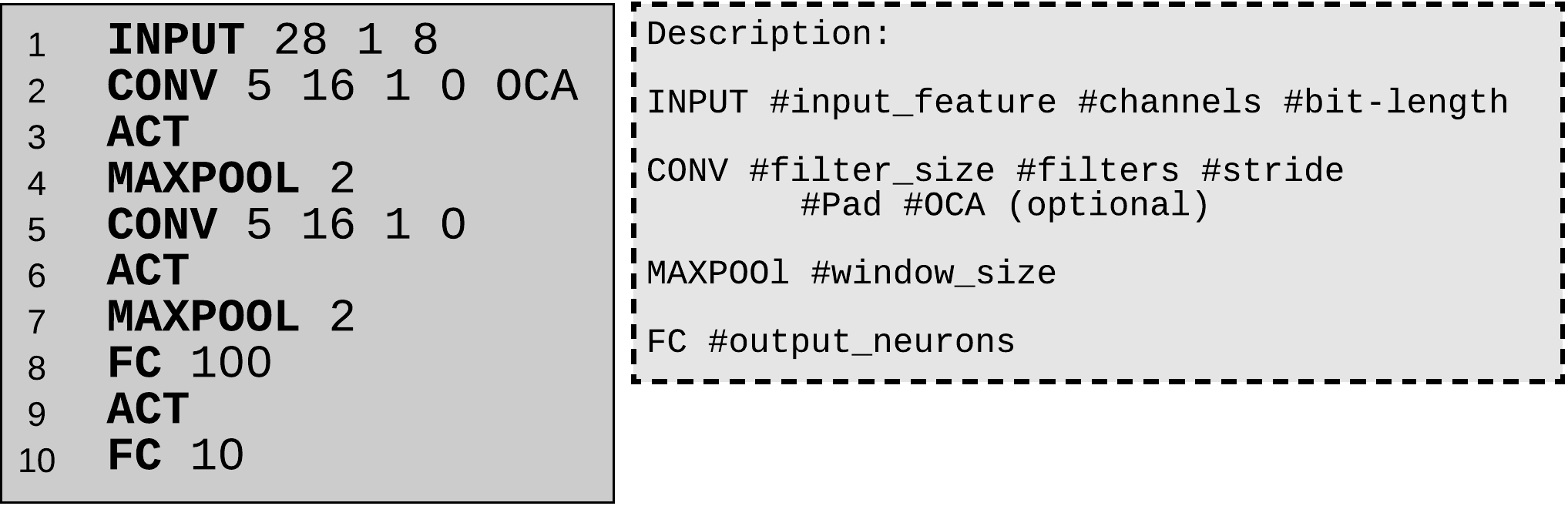}
    \vspace{-0.2em}
    \caption{Sample snippet code in \sys{}.}
    \label{fig:code}
\end{figure}

\sys{} provides a simplified and easy-to-use API for oblivious inference. The framework accepts a high-level description of the network, parameters of each layer, and input structure. 
It automatically computes the number of invocations and the interconnection between all of the base circuits. \fig{fig:code} shows the complete network description that a user needs to write for a sample network architecture (the BM3 architecture, see \sect{sec:eval}). 
All of the required circuits are automatically generated using TinyGarble~\cite{songhori2015tinygarble} synthesis libraries.
It is worth mentioning that for the task of oblivious inference, our API is much simpler compared to the recent {\it high-level} EzPC framework~\cite{ezpc}. For example, the required lines of code to describe BM1, BM2, and BM3 network architectures (see \sect{sec:eval}) in EzPC are 78, 88, and 154, respectively. In contrast, they can be described with only 6, 6, and 10 lines of code in our framework. 

\noindent {\bf Keras to \sys{} Translation.} To further facilitate the adaptation of \sys{}, a compiler is created to translate the description of the neural network in Keras~\cite{chollet2015keras} to the \sys{} format. The compiler creates the \texttt{.xonn} file and puts the network parameters into the required format (\texttt{HEX} string) to be read by the framework during the execution of the GC protocol. All of the parameter adjustments are also automatically performed by the compiler. 
\section{Related Work}\label{sec:rw}
\vspace{-0.2cm}
One of the early solutions for oblivious inference is proposed by Barni et al.~\cite{barni2006privacy} that leverages homomorphic encryption. 
Howerver, this work leaks some information about the intermediate states of the computation.  
Orlandi et al.~\cite{orlandi2007oblivious} improve upon this work and avoid any information leakage. 
In~\cite{sadeghi2008generalized}, authors propose a mechanism to also keep the topology of the neural network secret, at the cost of more computation overhead.
A solution based on additively homomorphic encryption and garbled circuits is presented in~\cite{barni2011privacy}.

CryptoNets~\cite{cryptonets} suggests the adaptation of Leveled Homomorphic Encryption (LHE) to perform oblivious inference. 
LHE is a variant of Partially HE that enables evaluation of depth-bounded arithmetic circuits.  
DeepSecure~\cite{deepsecure} is a privacy-preserving DL framework that relies on the GC protocol. 
CryptoDL~\cite{cryptodl} improves upon CryptoNets~\cite{cryptonets} and proposes more efficient approximation of the non-linear functions using low-degree polynomials. Their solution is based on LHE and uses mean-pooling in replacement of the max-pooling layer.
Chou et al. propose to utilize the sparsity within the DL model to accelerate the inference~\cite{chou2018faster}.

SecureML~\cite{secureml} is a privacy-preserving machine learning framework based on homomorphic encryption, GC, and secret sharing. SecureML also uses customized activation functions and supports privacy-preserving training in addition to inference. Two non-colluding servers are used to train the DL model where each client XOR-shares her input and sends the shares to both servers. 
MiniONN~\cite{minionn} is a mixed-protocol framework for oblivious inference. The underlying cryptographic protocols are HE, GC, and secret sharing. 

Chameleon~\cite{chameleon} is a more recent mixed-protocol framework for machine learning, i.e., Support Vector Machines (SVMs) as well as DNNs.
Authors propose to perform low-depth non-linear functions using the Goldreich-Micali-Wigderson (GMW) protocol~\cite{goldreich1987play}, high-depth functions by the GC protocol, and linear operations using additive secret sharing. Moreover, they propose to use correlated randomness to more efficiently compute linear operations. 
EzPC~\cite{ezpc} is a secure computation framework that enables users to write high-level programs and create a protocol based on both Boolean and Arithmetic shares.
The back-end cryptographic engine is based on the ABY framework~\cite{aby}.
HyCC~\cite{buscher2018hycc} is a high-level secure computation framework that complies ANSI C programs into efficient hybrid protocols. 

Shokri and Shmatikov~\cite{shokri2015privacy} proposed a solution for privacy-preserving collaborative deep learning where the training data is distributed among many parties. Their approach, enables clients to train their local model on their own training data and update the central model's parameters held by a central server. However, it has been shown that a malicious client can learn significant information about the other client's private data~\cite{hitaj2017deep}. 
Google~\cite{bonawitz2017practical} has recently introduced a new approach \rebut{for securely aggregating} the parameter updates from multiple users. 
However, none of these approaches~\cite{shokri2015privacy, bonawitz2017practical} study the oblivious inference problem. An overview of related frameworks is provided in~\cite{survey1,survey2}.

Frameworks such as $\text{ABY}^3$~\cite{mohassel2018aby} and SecureNN~\cite{waghsecurenn} have different computation models and they rely on three (or four) parties during the oblivious inference. In contrast, \sys{} does not require an additional server for the computation. 
In E2DM framework~\cite{jiang2018secure}, the model owner can encrypt and outsource the model to an untrusted server to perform oblivious inference.
Concurrently and independently of ours, in TAPAS~\cite{sanyal2018tapas}, Sanyal et al. study the binarization of neural networks in the context of oblivious inference. 
They report inference latency of 147 seconds on MNIST dataset with 98.6\% prediction accuracy using custom CNN architecture.
However, as we show in \sect{sec:eval} (BM3 benchmark), \sys{} outperforms TAPAS by close to {\it three orders of magnitude}. 

\gazelle{}~\cite{gazelle} is the previously most efficient oblivious inference framework. It is a mixed-protocol approach based on additive HE and GC. In Gazelle, convolution operations are performed using the packing property of HE. In this approach, many numbers are packed inside a single ciphertext for faster convolutions. 
In \sect{sec:circ}, we briefly discuss one of the essential requirements that the \gazelle{} protocol has to satisfy in order to be secure, namely, {\it circuit privacy}. 

\vspace{0.2em}
\noindent {\bf High-Level Comparison.} In contrast to prior art, we propose a DL-secure computation co-design approach.
To the best of our knowledge, DeepSecure~\cite{deepsecure} is the only solution that preprocesses the data and network before the secure computation protocol. However, the preprocessing step reveals some information about the network parameters and structure of data.
Moreover, another advantage of \sys{} is the constant round complexity regardless of the number of layers in the NN.
It has been shown that round complexity is one of the important criteria in designing secure computation protocols~\cite{ben2016optimizing} since the performance can significantly be reduced where the network latency is high.
As we show in \sect{sec:eval}, \sys{} outperforms all previous solutions in inference latency.
\tab{tab:high} summarizes a high-level comparison between state-of-the-art oblivious inference frameworks. 

\vspace{-0.2em}
\begin{table}[ht]
\centering
\caption{High-Level Comparison of oblivious inference frameworks. ``{\bf C}''onstant round complexity. ``{\bf D}''eep learning/secure computation co-design. ``{\bf I}''ndependence of secondary server. ``{\bf U}''pgradeable to malicious security using standard solutions. ``{\bf S}''upporting any non-linear layer.}
\vspace{-0.6em}
\label{tab:high}
\scalebox{0.85}{
\begin{tabular}{lllcccc} \toprule
\textbf{Framework}          				& \textbf{Crypto. Protocol}  & \textbf{C} &\textbf{D}& \textbf{I} & \textbf{U} & \textbf{S} \\\midrule
CryptoNets~\cite{cryptonets} 	& HE                  	& \cmark         & \xmark & \cmark  & \xmark & \xmark \\
DeepSecure~\cite{deepsecure} 	& GC                    		& \cmark         & \cmark & \cmark  & \cmark & \cmark \\
SecureML~\cite{secureml}   		& HE, GC, SS                   	& \xmark           & \xmark & \xmark  & \xmark & \xmark \\
MiniONN~\cite{minionn}    		& HE, GC, SS         & \xmark           & \xmark & \cmark  & \xmark & \cmark \\
Chameleon~\cite{chameleon} 		& GC, GMW, SS 			& \xmark 		   & \xmark & \xmark  & \xmark & \cmark \\
EzPC~\cite{ezpc}       			& GC, SS               & \xmark           & \xmark & \cmark  & \xmark & \cmark \\
\gazelle{}~\cite{gazelle}       	& HE, GC, SS          & \xmark           & \xmark & \cmark  & \xmark & \cmark \\
{\bf \sys{} (This work)} 				& GC, SS               & \cmark         & \cmark & \cmark  & \cmark & \cmark \\
\bottomrule
\end{tabular}
}
\end{table}

\section{Circuit Privacy}\label{sec:circ} 
In \gazelle{}~\cite{gazelle}, for each linear layer, the protocol starts with a vector $\mathbf{m}$ that is secret-shared between client $\mathbf{m_1}$ and server $\mathbf{m_2}$ ($\mathbf{m} = \mathbf{m_1} + \mathbf{m_2}$).
The protocol outputs the secret shares of the vector $\mathbf{m'} = A\cdot \mathbf{m}$ where $A$ is a matrix known to the server but not to the client. 
The protocol has the following procedure: (i) Client generates a pair $(pk, sk)$ of public and secret keys of an additive homomorphic encryption scheme \textsf{HE}. (ii) Client sends \textsf{HE}$.Enc_{pk}(\mathbf{m_1})$ to the server. Server adds its share ($\mathbf{m_2}$) to the ciphertext and recovers encryption of $\mathbf{m}$: \textsf{HE}$.Enc_{pk}(\mathbf{m})$. (iii) Server homomorphically evaluates the multiplication with $A$ and obtains the encryption of $\mathbf{m'}$. (iv) Server secret shares $\mathbf{m'}$ by sampling a random vector $\mathbf{r}$ and returns ciphertext $\mathbf{c}=$\textsf{HE}$.Enc_{pk}(\mathbf{m'-r})$ to the client. 
The client can decrypt $\mathbf{c}$ using private key $sk$ and obtain $\mathbf{m' - r}$. 

\gazelle{} uses the Brakerski-Fan-Vercauteren (BFV) scheme~\cite{brakerski2012fully,fan2012somewhat}. 
However, the vanilla BFV scheme does not provide circuit privacy. At high-level, the circuit privacy requirement states that the ciphertext $\mathbf{c}$ should not reveal any information about the private inputs to the client (i.e., $A$ and $\mathbf{r}$) other than the underlying plaintext $A\cdot \mathbf{m} - \mathbf{r}$. Otherwise, some information is leaked. \gazelle{} proposes two methods to provide circuit privacy that are not incorporated in their implementation. Hence, we need to scale up their performance numbers for a fair comparison.

The first method is to let the client and server engage in a two-party secure decryption protocol, where the input of client is $sk$ and input of server is $\mathbf{c}$. However, this method adds communication and needs extra rounds of interaction. A more widely used approach is {\it noise flooding}. Roughly speaking, the server adds a large noise term to $\mathbf{c}$ before returning it to the client. The noise is big enough to drown any extra information contained in the ciphertext, and still small enough to so that it still decrypts to the same plaintext. 

For the concrete instantiation of \gazelle{}, one needs to triple the size of ciphertext modulus $q$ from 60 bits to 180 bits, and increase the ring dimension $n$ from 2048 to 8192. The (amortized) complexity of homomorphic operations in the BFV scheme is approximately $O( \log n\ \log q)$, with the exception that some operations run in $O(\log q)$ amortized time. Therefore, adding noise flooding would result in a {\it 3-3.6 times slow down} for the HE component of \gazelle{}. To give some concrete examples, we consider two networks used for benchmarking in \gazelle{}: MNIST-D and CIFAR-10 networks. For the MNIST-D network, homomorphic encryption takes 55\% and 22\% in online and total time, respectively. For CIFAR-10, the corresponding figures are 35\%, and 10\%\footnote{these percentage numbers are obtained through private communication with the authors.}. Therefore, we estimate that the total time for MNIST-D will grow from 0.81s to 1.16-1.27s (network BM3 in this paper). In the case of CIFAR-10 network, the total time will grow from 12.9s to 15.48-16.25s.

\vspace{-0.8em}
\section{Experimental Results}\label{sec:eval}

We evaluate \sys{} on MNIST and CIFAR10 datasets, which are two popular classification benchmarks used in prior work.
\rebut{In addition, we provide four healthcare datasets to illustrate the applicability of \sys{} in real-world scenarios.}
For training \sys{}, we use Keras~\cite{chollet2015keras} with Tensorflow backend~\cite{tensorflow}. \rebut{The source code of \sys{} is compiled with GCC 5.5.0 using O3 optimization. All Boolean circuits are synthesized using Synopsys Design Compiler 2015.} Evaluations are performed on (Ubuntu 16.04 LTS) machines with Intel-Core i7-7700k and $32$GB of RAM. \rebut{The experimental setup is comparable (but has less computational power) compared to the prior art~\cite{gazelle}. Consistent with prior frameworks, we evaluate the benchmarks in the LAN setting.} 



\subsection{Evaluation on MNIST}
There are mainly three network architectures that prior works have implemented for the MNIST dataset. We convert these reference networks into their binary counterparts and train them using the standard BNN training algorithm~\cite{courbariaux2016binarized}.
\tab{tab:mnist_archs} summarizes the architectures for the MNIST dataset.


\begin{table}[h]
\centering
\caption{Summary of the trained binary network architectures evaluated on the MNIST dataset. Detailed descriptions are available in \app{sec:runtime_app}, Table~\ref{tab:archs_details}.}
\label{tab:mnist_archs}
\resizebox{\columnwidth}{!}{
\begin{tabular}{ccc}\toprule   
\textbf{Arch.} & \textbf{Previous Papers}                                                                              & \textbf{Description}                    \\ \midrule
BM1                   & SecureML~\cite{secureml}, MiniONN~\cite{minionn}                                                                                     & 3 \lname{FC}                            \\ 
\multirow{2}{*}{BM2}  & \multirow{2}{*}{\begin{tabular}[c]{@{}c@{}}CryptoNets~\cite{cryptonets}, MiniONN~\cite{minionn},\\ DeepSecure~\cite{deepsecure}, Chameleon~\cite{chameleon}\end{tabular}} & \multirow{2}{*}{1 \lname{CONV}, 2 \lname{FC}} \\
                      &                                                                                                       &                                         \\ 
BM3                   & MiniONN~\cite{minionn}, EzPC~\cite{ezpc}                                                                                         & 2 \lname{CONV}, 2\lname{MP}, 2\lname{FC} \\\bottomrule                
\end{tabular}
}
\end{table}

\begin{figure}[ht!]
    \centering
    \begin{subfigure}[b]{0.48\columnwidth}
            \includegraphics[width=\columnwidth]{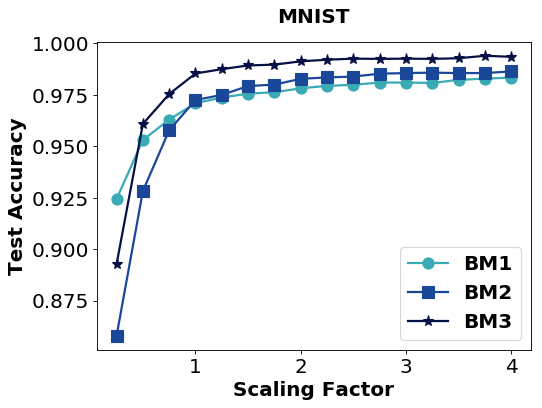}
            \caption{}\label{fig:acc_widening_mnist}
    \end{subfigure}
    ~
    \centering
    \begin{subfigure}[b]{0.46\columnwidth}
            \includegraphics[width=\columnwidth]{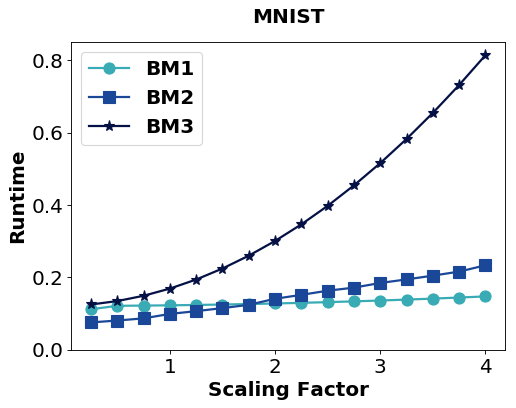}
            \caption{}\label{fig:time_widening_mnist}
    \end{subfigure}
    \caption{Effect of scaling factor on (a)~accuracy and (b)~inference runtime of MNIST networks.  No pruning was applied in this evaluation.}
    \label{fig:widening_mnist}
\end{figure}

\vspace{0.2em}
\noindent{\bf Analysis of Network Scaling:} Recall that the classification accuracy of \sys{} is controlled by scaling the number of neurons in all layers (Section~\ref{ssec:binarization}). Figure~\ref{fig:acc_widening_mnist} depicts the inference accuracy with different scaling factors (more details in Table~\ref{tab:mnist_details} in \app{sec:runtime_app}). As we increase the scaling factor, the accuracy of the network increases. This accuracy improvement comes at the cost of a higher computational complexity of the (scaled) network. As a result, increasing the scaling factor leads to a higher runtime. Figure~\ref{fig:time_widening_mnist} depicts the runtime of different BNN architectures as a function of the scaling factor $s$. 
\rebut{Note that the runtime grows (almost) quadratically with the scaling factor due to the quadratic increase in the number of $Popcount$ operations in the neural network (see $BM3$). However, for the $BM1$ and $BM2$ networks, the overall runtime is dominated by the constant initialization cost of the OT protocol ($\sim70$ millisecond).}


\begin{figure}[h]
 \centering
\begin{subfigure}[b]{0.49\columnwidth}
\includegraphics[width=\columnwidth]{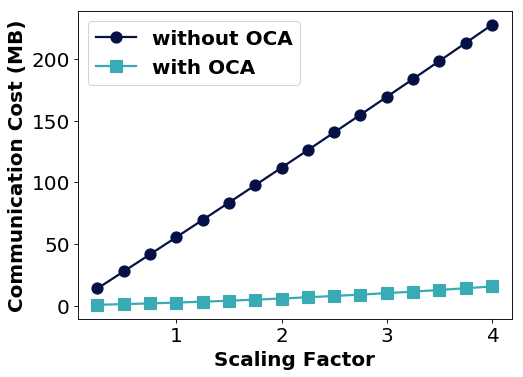}
\end{subfigure}
~
\begin{subfigure}[b]{0.48\columnwidth}
\includegraphics[width=\columnwidth]{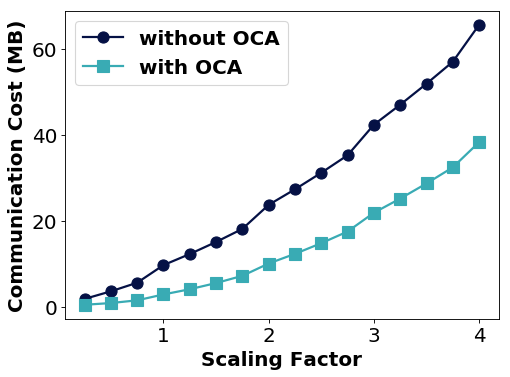}
\end{subfigure}
\caption{Effect of OCA on the communication of the BM1 (left) and BM2 (right) networks for different scaling factors. No pruning was applied in this evaluation. }
\label{fig:comm_widening_mnist}
\end{figure}

\noindent{\bf GC Cost and the Effect of OCA:} The communication cost of GC is the key contributor to the overall runtime of \sys{}. Here, we analyze the effect of the scaling factor on the total message size. Figure~\ref{fig:comm_widening_mnist} shows the communication cost of GC for the BM1 and BM2 network architectures. As can be seen, the message size increases with the scaling factor.
We also observe that the OCA protocol drastically reduces the message size. This is due to the fact that the first layer of BM1 and BM2 models account for a large portion of the overall computation; hence, improving the first layer with OCA has a drastic effect on the overall communication.



\vspace{0.3em}
\noindent{\bf Comparison to Prior Art:} We emphasize that, unlike previous work, the accuracy of \sys{} can be customized by tuning the scaling factor ($s$). Furthermore, our channel/neuron pruning step (Algorithm~\ref{alg:iterative_pruning}) can reduce the GC cost in a post-processing phase. To provide a fair comparison between \sys{} and prior art, we choose a proper scaling factor and trim the pertinent scaled BNN such that the corresponding BNN achieves the same accuracy as the previous work. Table~\ref{tab:comparison_mnist} compares \sys{} with the previous work in terms of accuracy, latency, and communication cost (a.k.a., message size). The last column shows the scaling factor ($s$) used to increase the width of the hidden layers of the BNN. Note that the scaled network is further trimmed using Algorithm~\ref{alg:iterative_pruning}. 

In \sys{}, the runtime for oblivious transfer is at least $\sim0.07$ second for initiating the protocol and then it grows linearly with the size of the garbled tables; As a result, in very small architectures such as $BM1$, our solution is slightly slower than previous works since the constant runtime dominates the total runtime. 
\rebut{However, for the $BM3$ network which has higher complexity than $BM1$ and $BM2$, \sys{} achieves a more prominent advantage over prior art.}
In summary, our solution achieves up to $7.7\times$ faster inference (average of $3.4\times$) compared to \gazelle{}~\cite{gazelle}.
Compared to MiniONN~\cite{minionn}, \sys{} has up to $62\times$ lower latency (average of $26\times$) \tab{tab:comparison_mnist}. Compared to EzPC~\cite{ezpc}, our framework is $34\times$ faster. \sys{} achieves $37.5\times$, $1859\times$, $60.4\times$, and $14\times$ better latency compared to SecureML~\cite{secureml}, CryptoNets~\cite{cryptonets}, DeepSecure~\cite{deepsecure}, and Chameleon~\cite{chameleon}, respectively.

\begin{table}[h]
\centering
\caption{\rebut{Comparison of \sys{} with the state-of-the-art for the MNIST network architectures.}}
\label{tab:comparison_mnist}
\resizebox{\columnwidth}{!}{
\begin{tabular}{cccccc}
\hline
{\bf Arch}.                & {\bf Framework}  & \multicolumn{1}{l}{{\bf Runtime (s)}} & \multicolumn{1}{l}{{\bf Comm. (MB)}} & {\bf Acc.} (\%) & {\bf s}    \\ \hline
\multirow{4}{*}{BM1} & SecureML   & 4.88                            & -                                      & 93.1      & -    \\
                     & MiniONN    & 1.04                            & 15.8                                   & 97.6      & -    \\
                     & \rebut{EzPC}    & \rebut{0.7}                           & \rebut{76}                                   & \rebut{97.6}      & \rebut{-}    \\
                     & \gazelle{}    & 0.09                            & 0.5                                    & 97.6      & -    \\
                     & \sys{}       & 0.13                            & 4.29                                   & 97.6      & 1.75 \\ \hline
\multirow{6}{*}{BM2} & CryptoNets & 297.5                           & 372.2                                  & 98.95     & -    \\
                     & DeepSecure & 9.67                            & 791                                    & 98.95     & -    \\
                     & MiniONN    & 1.28                            & 47.6                                   & 98.95     & -    \\
                     & Chameleon  & 2.24                            & 10.5                                   & 99.0      & -    \\
                     & \rebut{EzPC}  & \rebut{0.6}                            & \rebut{70}                                  & \rebut{99.0}      & \rebut{-}    \\
                     & \gazelle{}    & 0.29                            & 8.0                                    & 99.0      & -    \\
                     & \sys{}       & 0.16                            & 38.28                                  & 98.64     & 4.00 \\ \hline
\multirow{4}{*}{BM3} & MiniONN    & 9.32                            & 657.5                                  & 99.0      & -    \\
                     & EzPC       & 5.1                             & 501                                    & 99.0      & -    \\
                     & \gazelle{}    & 1.16                            & 70                                     & 99.0      & -    \\
                     & \sys{}       & 0.15                            & 32.13                                  & 99.0      & 2.00 \\ \hline
\end{tabular}
}
\end{table}

\subsection{Evaluation on CIFAR-10}
In Table~\ref{tab:cifar_archs}, we summarize the network architectures that we use for the CIFAR-10 dataset. In this table, BC1 is the binarized version of the architecture proposed by MiniONN. To evaluate the scalability of our framework to larger networks, we also binarize the Fitnet~\cite{romero2014fitnets} architectures, which are denoted as BC2-BC5. We also evaluate \sys{} on the popular VGG16 network architecture (BC6). 
Detailed architecture descriptions are available in \app{sec:runtime_app}, Table~\ref{tab:archs_details}.
\vspace{+0.15cm}

\begin{table}[h]
\centering
\caption{Summary of the trained binary network architectures evaluated on the CIFAR-10 dataset. 
}
\label{tab:cifar_archs}
\resizebox{0.99\columnwidth}{!}{
\begin{tabular}{ccc}
\hline
\textbf{Arch.} & \textbf{Previous Papers}                                                    & \textbf{Description} \\ \hline
BC1            & \begin{tabular}[c]{@{}c@{}}MiniONN\cite{minionn}, Chameleon~\cite{chameleon},\\ EzPC~\cite{ezpc}, \gazelle{}~\cite{gazelle}\end{tabular} & 7 \lname{CONV}, 2 \lname{MP}, 1 \lname{FC}    \\ 
BC2            & Fitnet~\cite{romero2014fitnets}                                                                     & 9 \lname{CONV}, 3 \lname{MP}, 1 \lname{FC}    \\ 
BC3            & Fitnet~\cite{romero2014fitnets}                                                                    & 9 \lname{CONV}, 3 \lname{MP}, 1 \lname{FC}    \\ 
BC4            & Fitnet~\cite{romero2014fitnets}                                                                     & 11 \lname{CONV}, 3 \lname{MP}, 1 \lname{FC}   \\ 
BC5            & Fitnet~\cite{romero2014fitnets}                                                                    & 17 \lname{CONV}, 3 \lname{MP}, 1 \lname{FC}   \\ 
BC6            & VGG16~\cite{simonyan2014very}                                                                      & 13 \lname{CONV},  5 \lname{MP}, 3 \lname{FC} \\ \hline
\end{tabular}
}
\end{table}

\begin{figure}[ht!]
    \centering
    \begin{subfigure}[b]{0.48\columnwidth}
            \includegraphics[width=\columnwidth]{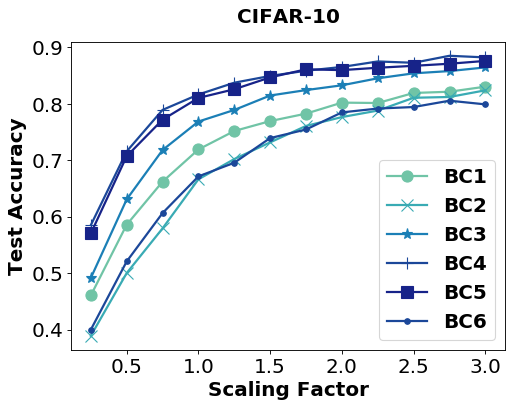}
            \caption{}\label{fig:acc_widening_cifar}
    \end{subfigure}
    ~
    \centering
    \begin{subfigure}[b]{0.48\columnwidth}
            \includegraphics[width=\columnwidth]{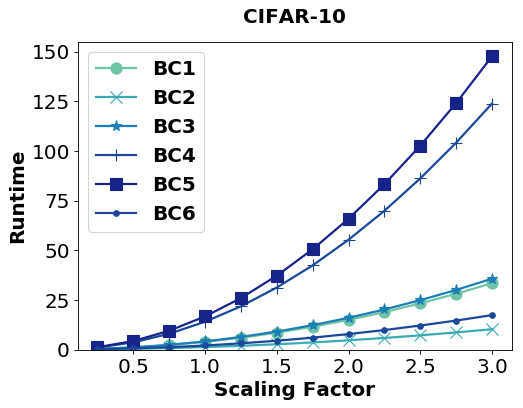}
            \caption{}\label{fig:time_widening_cifar}
    \end{subfigure}
    ~
    \caption{(a)~Effect of scaling factor on accuracy for CIFAR-10 networks. 
    (b)~Effect of scaling factor on runtime. No pruning was applied in this evaluation.
    }
    \label{fig:widening_cifar}
\end{figure}

\vspace{+0.15cm}
\noindent{\bf Analysis of Network Scaling:} Similar to the analysis on the MNIST dataset, we show that the accuracy of our binary models for CIFAR-10 can be tuned based on the scaling factor that determines the number of neurons in each layer. Figure~\ref{fig:acc_widening_cifar} depicts the accuracy of the BNNs with different scaling factors. As can be seen, increasing the scaling factor enhances the classification accuracy of the BNN. The runtime also increases with the scaling factor as shown in Figure~\ref{fig:time_widening_cifar} (more details in Table~\ref{tab:cifar_details}, \app{sec:runtime_app}). 


\noindent{\bf Comparison to Prior Art:} We scale the BC2 network with a factor of $s=3$, then prune it using Algorithm~\ref{alg:iterative_pruning}. Details of pruning steps are available in Table~\ref{tab:trimming_cifar} in \app{sec:trimming_eval}. The resulting network is compared against prior art in Table~\ref{tab:comparison_cifar}. 
As can be seen, our solution achieves $2.7\times$, $45.8\times$, $9.1\times$, and $93.1\times$ lower latency compared to \gazelle{}, EzPC, Chameleon, and MiniONN, respectively.


\begin{table}[h]
\centering
\caption{Comparison of \sys{} \rebut{with prior art on CIFAR-10.}}
\vspace{-0.5em}
\label{tab:comparison_cifar}
\resizebox{0.9\columnwidth}{!}{
\begin{tabular}{ccccc}
\hline
{\bf Framework} & {\bf Runtime (s)} & {\bf Comm. (MB)} & {\bf Acc. (\%)} & {\bf s}    \\ \hline
MiniONN   & 544         & 9272               & 81.61     & -    \\
Chameleon & 52.67       & 2650               & 81.61     & -    \\
EzPC      & 265.6       & 40683              & 81.61     & -    \\
\gazelle{}   & 15.48       & 1236               & 81.61     & -    \\
\sys{}      & 5.79        & 2599               & 81.85     & 3.00 \\ \hline
\end{tabular}
\vspace{-0.2cm}
}
\end{table}

\vspace{-0.5cm}
\subsection{Evaluation on Medical Datasets}
One of the most important applications of oblivious inference is medical data analysis. Recent advances in deep learning greatly benefit many complex diagnosis tasks that require exhaustive manual inspection by human experts~\cite{esteva2019guide,esteva2017dermatologist,alipanahi2015predicting,rajkomar2018scalable}. To showcase the applicability of oblivious inference in real-world medical applications, we provide several benchmarks for publicly available healthcare datasets summarized in Table~\ref{tab:medical}. We split the datasets into validation and training portions as indicated in the last two columns of Table~\ref{tab:medical}.  
All datasets except Malaria Infection are normalized to have $0$ mean and standard deviation of $1$ per feature. The images of Malaria Infection dataset are resized to $32\times32$ pictures. The normalized datasets are quantized up to 3 decimal digits. Detailed architectures are available in~\app{sec:runtime_app}, Table~\ref{tab:archs_details}
We report the validation accuracy along with inference time and message size in Table~\ref{tab:medical_evaluations}.

\begin{table}[h]
\caption{Summary of medical application benchmarks.}\label{tab:medical}
\resizebox{\columnwidth}{!}{
\begin{tabular}{ccccc}
\hline
\multirow{2}{*}{{\bf Task}} & \multirow{2}{*}{{\bf Arch.}} & \multirow{2}{*}{{\bf Description}} & \multicolumn{2}{c}{{\bf \# of Samples}} \\ \cline{4-5} 
 &  &  & Tr. & Val. \\ \hline
Breast Cancer~\cite{breastcancer} & BH1 & 3 \lname{FC} & 453 & 113 \\
Diabetes~\cite{diabetes} & BH2 & 3 \lname{FC} & 615 & 153 \\
Liver Disease~\cite{liver} & BH3 & 3 \lname{FC} & 467 & 116 \\
Malaria Infection~\cite{malaria} & BH4 & \begin{tabular}[c]{@{}c@{}}2 \lname{CONV},\\ 2 \lname{MP}, 2 \lname{FC}\end{tabular} & 24804 & 2756 \\
\hline
\end{tabular}
}
\end{table}

\begin{table}[h]
\centering
\caption{Runtime, communication cost (Comm.), and accuracy (Acc.) for medical benchmarks.}\label{tab:medical_evaluations}
\resizebox{0.83\columnwidth}{!}{
\begin{tabular}{cccc}
\hline
\textbf{Arch.} & \textbf{Runtime (ms)} & \textbf{Comm. (MB)} & \textbf{Acc. (\%)} \\ \hline
BH1 & 82 & 0.35 & 97.35  \\
BH2 & 75 & 0.16 & 80.39  \\
BH3 & 81 & 0.3  & 80.17  \\
BH4 & 482 & 120.75 & 95.03  \\ \hline
\end{tabular}
}
\end{table}

\vspace{-0.9em}
\section{Conclusion}\label{sec:conc}
\vspace{-0.4em}
We introduce \sys{}, a novel framework to automatically train and use deep neural networks for the task of oblivious inference. \sys{} utilizes Yao's Garbled Circuits (GC) protocol and relies on binarizing the DL models in order to translate costly matrix multiplications to \texttt{XNOR} operations that are free in the GC protocol. Compared to \gazelle{}~\cite{gazelle}, prior best solution, \sys{} achieves $7\times$ lower latency.
Moreover, in contrast to \gazelle{} that requires one round of interaction for each layer, our solution needs a constant round of interactions regardless of the number of layers. 
Maintaining constant round complexity is an important requirement in Internet settings as a typical network latency can significantly degrade the performance of oblivious inference. 
Moreover, since our solution relies on the GC protocol, it can provide much stronger security guarantees such as security against malicious adversaries using standard cut-and-choose protocols. 
\sys{} high-level API enables clients to utilize the framework with a minimal number of lines of code. To further facilitate the adaptation of our framework, we design a compiler to translate the neural network description in Keras format to that of \sys{}.

\ifanonymous\else
\paragraph{Acknowledgements}
We would like to thank the anonymous reviewers for their insightful comments. 
\fi

\bibliographystyle{plain}
\bibliography{main_bib}

\appendix

\section{Experimental Details}
\vspace{-0.25cm}
\subsection{Network Trimming Examples}\label{sec:trimming_eval}
 Table~\ref{tab:trimming_mnist} and~\ref{tab:trimming_cifar} summarize the trimming steps for the MNIST and CIFAR-10 benchmarks, respectively.
 \vspace{-0.2cm}
 

\begin{table}[h]
\caption{Trimming MNIST architectures.}\label{tab:trimming_mnist}
\vspace{-0.2cm}
\resizebox{0.95\columnwidth}{!}{
\begin{tabular}{ccccccc}
\hline
\multirow{2}{*}{Network}                                                & \multirow{2}{*}{Property} & \multicolumn{4}{c}{Trimming Step}   & \multirow{2}{*}{Change} \\ \cline{3-6}
                                                                        &                           & initial & step 1 & step 2 & step 3 &                         \\ \hline
\multirow{3}{*}{\begin{tabular}[c]{@{}c@{}}BM1\\ (s=1.75)\end{tabular}} & Acc. (\%)                 & 97.63   & 97.59  & 97.28  & 97.02  & -0.61\%                 \\
                                                                        & Comm. (MB)                 & 4.95    & 4.29   & 3.81   & 3.32   & 1.49$\times$ less              \\
                                                                        & Lat. (ms)                 & 158     & 131    & 114    & 102    & 1.54$\times$ faster            \\ \hline
\multirow{3}{*}{\begin{tabular}[c]{@{}c@{}}BM2\\ (s=4)\end{tabular}}    & Acc. (\%)                 & 98.64   & 98.44  & 98.37  & 98.13  & -0.51\%                 \\
                                                                        & Comm. (MB)                 & 38.28   & 28.63  & 24.33  & 15.76  & 2.42$\times$ less              \\
                                                                        & Lat. (ms)                 & 158     & 144    & 134    & 104    & 1.51$\times$ faster            \\ \hline
\multirow{3}{*}{\begin{tabular}[c]{@{}c@{}}BM3\\ (s=2)\end{tabular}}    & Acc. (\%)                 & 99.22   & 99.11  & 98.96  & 99.00  & -0.22\%                 \\
                                                                        & Comm. (MB)                 & 56.08   & 42.51  & 37.34  & 32.13  & 1.75$\times$ less              \\
                                                                        & Lat. (ms)                 & 190     & 165    & 157    & 146    & 1.3$\times$ faster             \\ \hline
\end{tabular}
}
\end{table}

\begin{table}[h]
\centering
\caption{Trimming the BC2 network for CIFAR-10.}\label{tab:trimming_cifar}
\vspace{-0.2cm}
\resizebox{0.85\columnwidth}{!}{
\begin{tabular}{cccccc}
\hline
\multirow{2}{*}{Property} & \multicolumn{4}{c}{Trimming Step}   & \multirow{2}{*}{Change} \\ \cline{2-5}
                          & initial & step 1 & step 2 & step 3 &                         \\ \hline
Acc. (\%)                 & 82.40   & 82.39  & 82.41  & 81.85  &   -0.55\%                      \\
Com. (GB)                 & 3.38    & 3.05   & 2.76   & 2.60   &1.30$\times$ less                         \\
Lat. (s)                 & 7.59    &6.87    & 6.23       & 5.79       & 1.31$\times$ faster                        \\ \hline
\end{tabular}
}
\end{table}

\newpage
\subsection{Accuracy, Runtime, and Communication}\label{sec:runtime_app}

\renewcommand{\arraystretch}{0.9}
\rebut{Runtime and communication reports are available in \tab{tab:mnist_details} and \tab{tab:cifar_details} for MNIST and CIFAR-10 benchmarks, respectively. The corresponding neural network architectures are provided in~\tab{tab:archs_details}. Entries corresponding to a communication of more than $40$GB are estimated using numerical runtime models.}

\begin{table}[h]
\centering
\caption{Accuracy (Acc.), communication (Comm.), and latency (Lat.) for MNIST dataset. Channel/neuron trimming is not applied.}
\label{tab:mnist_details}
\resizebox{0.7\columnwidth}{!}{
\begin{tabular}{ccccc}
\toprule
Arch.                                     & s   & Acc. (\%) & Comm. (MB) & Lat. (s) \\ \hline
\multicolumn{1}{c}{\multirow{5}{*}{BM1}}&1&97.10&2.57&0.12\\
\multicolumn{1}{c}{} & 1.5&97.56&4.09&0.13\\ 
\multicolumn{1}{c}{} & 2&97.82&5.87&0.13\\ 
\multicolumn{1}{c}{} & 3&98.10&10.22&0.14\\ 
\multicolumn{1}{c}{} & 4&98.34&15.62&0.15\\ \hline
\multicolumn{1}{c}{\multirow{5}{*}{BM2}}&1&97.25&2.90&0.10\\
\multicolumn{1}{c}{} &1.50&97.93&5.55&0.12\\
\multicolumn{1}{c}{} &2&98.28&10.09&0.14\\
\multicolumn{1}{c}{} &3&98.56&21.90&0.18\\
\multicolumn{1}{c}{} &4&98.64&38.30&0.23\\ \hline
\multicolumn{1}{c}{\multirow{5}{*}{BM3}}& 1&98.54&17.59&0.17\\
\multicolumn{1}{c}{} & 1.5&98.93&36.72&0.22\\
\multicolumn{1}{c}{} & 2&99.13&62.77&0.3\\
\multicolumn{1}{c}{} & 3&99.26&135.88&0.52\\
\multicolumn{1}{c}{} & 4&99.35&236.78&0.81\\ \bottomrule

\end{tabular}}
\end{table}

\begin{table}[h]
\centering
\caption{Accuracy (Acc.), communication (Comm.), and latency (Lat.) for CIFAR-10 dataset. Channel/neuron trimming is not applied.}
\label{tab:cifar_details}
\resizebox{0.7\columnwidth}{!}{
\begin{tabular}{ccccc}
\toprule
Arch.                                     & s   & Acc. (\%) & Comm. (MB) & Lat. (s) \\ \hline
\multicolumn{1}{c}{\multirow{5}{*}{BC1}} &1&0.72&1.26&3.96\\
\multicolumn{1}{c}{} &1.5&0.77&2.82&8.59\\
\multicolumn{1}{c}{} &2&0.80&4.98&15.07\\
\multicolumn{1}{c}{} &3&0.83&11.15&33.49\\\hline
\multicolumn{1}{c}{\multirow{5}{*}{BC2}}&1&0.67&0.39&1.37\\
\multicolumn{1}{c}{} &1.5&0.73&0.86&2.78\\
\multicolumn{1}{c}{} &2&0.78&1.53&4.75\\
\multicolumn{1}{c}{} &3&0.82&3.40&10.35\\ \hline
\multicolumn{1}{c}{\multirow{5}{*}{BC3}} &1&0.77&1.35&4.23\\
\multicolumn{1}{c}{}& 1.5&0.81&3.00&9.17\\
\multicolumn{1}{c}{}& 2&0.83&5.32&16.09\\
\multicolumn{1}{c}{}& 3&0.86&11.89&35.77\\\hline
\multicolumn{1}{c}{\multirow{5}{*}{BC4}}&1&0.82&4.66&14.12\\
\multicolumn{1}{c}{}&1.5&0.85&10.41&31.33\\
\multicolumn{1}{c}{}&2&0.87&18.45&55.38\\
\multicolumn{1}{c}{}&3&0.88&41.37&123.94\\ \hline
\multicolumn{1}{c}{\multirow{5}{*}{BC5}} & 1&0.81&5.54&16.78 \\
\multicolumn{1}{c}{}&1.5&0.85&12.40&37.29\\
\multicolumn{1}{c}{}&2 &0.86&21.98&65.94\\
\multicolumn{1}{c}{}&3 &0.88&49.30&147.66\\ \hline
\multicolumn{1}{c}{\multirow{5}{*}{BC6}}&1&0.67&0.65&2.15\\
\multicolumn{1}{c}{}&1.5&0.74&1.46&4.55\\
\multicolumn{1}{c}{}&2&0.78&2.58&7.91\\
\multicolumn{1}{c}{}&3&0.80&5.77&17.44\\ \bottomrule

\end{tabular}}
\end{table}

\clearpage
\newpage
\newcolumntype{n}{>{\hsize=0.45\hsize}>{\footnotesize}X}
\newcolumntype{t}{>{\hsize=23\hsize}>{\footnotesize}X}
\renewcommand{\arraystretch}{0.9}
\begin{table}[h]
\caption{Evaluated network architectures.}\label{tab:archs_details}\vspace{-1em}
\centering
\resizebox{0.75\columnwidth}{!}{
\begin{tabular}{|ll|}
\multicolumn{2}{c}{\textbf{BM1}} \\\hline
1 &\textbf{FC} [input: $784$, output: $128s$] + \textbf{BN} + \textbf{BA}\\
2 &\textbf{FC} [input: $128s$, output: $128s$] + \textbf{BN} + \textbf{BA}\\
3 &\textbf{FC} [input: $128s$, output: $10$] + \textbf{BN} + \textbf{Softmax}\\\hline
\multicolumn{2}{c}{\textbf{BM2}} \\\hline
1 &\textbf{CONV} [input: $28 \times 28 \times 1$, window: $5 \times 5$, stride: $2$, kernels: $5s$, \\
&output: $12 \times 12 \times 5s$] + \textbf{BN} + \textbf{BA}\\
2 &\textbf{FC} [input: $720s$, output: $100s$] + \textbf{BN} + \textbf{BA}\\
3 &\textbf{FC} [input: $100s$, output: $10$] + \textbf{BN} + \textbf{Softmax}\\\hline
\multicolumn{2}{c}{\textbf{BM3}} \\\hline
1 &\textbf{CONV} [input: $28 \times 28 \times 1$, window: $5 \times 5$, stride: $1$, kernels: $16s$, \\
&output: $24 \times 24 \times 16s$] + \textbf{BN} + \textbf{BA}\\
2 &\textbf{MP} [input: $24 \times 24 \times 16s$, window: $2 \times 2$, output: $12 \times 12 \times 16s$]\\
3 &\textbf{CONV} [input: $12 \times 12 \times 16s$, window: $5 \times 5$, stride: $1$, kernels: $16s$, \\
&output: $8 \times 8 \times 16s$] + \textbf{BN} + \textbf{BA}\\
4 &\textbf{MP} [input: $8 \times 8 \times 16s$, window: $2 \times 2$, output: $4 \times 4 \times 16s$]\\
5 &\textbf{FC} [input: $256s$, output: $100s$] + \textbf{BN} + \textbf{BA}\\
6 &\textbf{FC} [input: $100s$, output: $10$] + \textbf{BN} + \textbf{Softmax}\\\hline
\multicolumn{2}{c}{\textbf{BC1}} \\\hline
1 &\textbf{CONV} [input: $32 \times 32 \times 3$, window: $3 \times 3$, stride: $1$, kernels: $64s$, \\
&output: $30 \times 30 \times 64s$] + \textbf{BN} + \textbf{BA}\\
2 &\textbf{CONV} [input: $30 \times 30 \times 64s$, window: $3 \times 3$, stride: $1$, kernels: $64s$, \\
&output: $28 \times 28 \times 64s$] + \textbf{BN} + \textbf{BA}\\
3 &\textbf{MP} [input: $28 \times 28 \times 64s$, window: $2 \times 2$, output: $14 \times 14 \times 64s$]\\
4 &\textbf{CONV} [input: $14 \times 14 \times 64s$, window: $3 \times 3$, stride: $1$, kernels: $64s$, \\
&output: $12 \times 12 \times 64s$] + \textbf{BN} + \textbf{BA}\\
5 &\textbf{CONV} [input: $12 \times 12 \times 64s$, window: $3 \times 3$, stride: $1$, kernels: $64s$, \\
&output: $10 \times 10 \times 64s$] + \textbf{BN} + \textbf{BA}\\
6 &\textbf{MP} [input: $10 \times 10 \times 64s$, window: $2 \times 2$, output: $5 \times 5 \times 64s$]\\
7 &\textbf{CONV} [input: $5 \times 5 \times 64s$, window: $3 \times 3$, stride: $1$, kernels: $64s$, \\
&output: $3 \times 3 \times 64s$] + \textbf{BN} + \textbf{BA}\\
8 &\textbf{CONV} [input: $3 \times 3 \times 64s$, window: $1 \times 1$, stride: $1$, kernels: $64s$, \\
&output: $3 \times 3 \times 64s$] + \textbf{BN} + \textbf{BA}\\
9 &\textbf{CONV} [input: $3 \times 3 \times 64s$, window: $1 \times 1$, stride: $1$, kernels: $16s$, \\
&output: $3 \times 3 \times 16s$] + \textbf{BN} + \textbf{BA}\\
10 &\textbf{FC} [input: $144s$, output: $10$] + \textbf{BN} + \textbf{Softmax}\\\hline
\multicolumn{2}{c}{\textbf{BC2}} \\\hline
1 &\textbf{CONV} [input: $32 \times 32 \times 3$, window: $3 \times 3$, stride: $1$, kernels: $16s$, \\
&output: $32 \times 32 \times 16s$] + \textbf{BN} + \textbf{BA}\\
2 &\textbf{CONV} [input: $32 \times 32 \times 16s$, window: $3 \times 3$, stride: $1$, kernels: $16s$, \\
&output: $32 \times 32 \times 16s$] + \textbf{BN} + \textbf{BA}\\
3 &\textbf{CONV} [input: $32 \times 32 \times 16s$, window: $3 \times 3$, stride: $1$, kernels: $16s$, \\
&output: $32 \times 32 \times 16s$] + \textbf{BN} + \textbf{BA}\\
4 &\textbf{MP} [input: $32 \times 32 \times 16s$, window: $2 \times 2$, output: $16 \times 16 \times 16s$]\\
5 &\textbf{CONV} [input: $16 \times 16 \times 16s$, window: $3 \times 3$, stride: $1$, kernels: $32s$, \\
&output: $16 \times 16 \times 32s$] + \textbf{BN} + \textbf{BA}\\
6 &\textbf{CONV} [input: $16 \times 16 \times 32s$, window: $3 \times 3$, stride: $1$, kernels: $32s$, \\
&output: $16 \times 16 \times 32s$] + \textbf{BN} + \textbf{BA}\\
7 &\textbf{CONV} [input: $16 \times 16 \times 32s$, window: $3 \times 3$, stride: $1$, kernels: $32s$, \\
&output: $16 \times 16 \times 32s$] + \textbf{BN} + \textbf{BA}\\
8 &\textbf{MP} [input: $16 \times 16 \times 32s$, window: $2 \times 2$, output: $8 \times 8 \times 32s$]\\
9 &\textbf{CONV} [input: $8 \times 8 \times 32s$, window: $3 \times 3$, stride: $1$, kernels: $48s$, \\
&output: $6 \times 6 \times 48s$] + \textbf{BN} + \textbf{BA}\\
10 &\textbf{CONV} [input: $6 \times 6 \times 48s$, window: $3 \times 3$, stride: $1$, kernels: $48s$, \\
&output: $4 \times 4 \times 48s$] + \textbf{BN} + \textbf{BA}\\
11 &\textbf{CONV} [input: $4 \times 4 \times 48s$, window: $3 \times 3$, stride: $1$, kernels: $64s$, \\
&output: $2 \times 2 \times 64s$] + \textbf{BN} + \textbf{BA}\\
12 &\textbf{MP} [input: $2 \times 2 \times 64s$, window: $2 \times 2$, output: $1 \times 1 \times 64s$]\\
13 &\textbf{FC} [input: $64s$, output: $10$] + \textbf{BN} + \textbf{Softmax}\\\hline
\multicolumn{2}{c}{\textbf{BC3}} \\\hline
1 &\textbf{CONV} [input: $32 \times 32 \times 3$, window: $3 \times 3$, stride: $1$, kernels: $16s$, \\
&output: $32 \times 32 \times 16s$] + \textbf{BN} + \textbf{BA}\\
2 &\textbf{CONV} [input: $32 \times 32 \times 16s$, window: $3 \times 3$, stride: $1$, kernels: $32s$, \\
&output: $32 \times 32 \times 32s$] + \textbf{BN} + \textbf{BA}\\
3 &\textbf{CONV} [input: $32 \times 32 \times 32s$, window: $3 \times 3$, stride: $1$, kernels: $32s$, \\
&output: $32 \times 32 \times 32s$] + \textbf{BN} + \textbf{BA}\\
4 &\textbf{MP} [input: $32 \times 32 \times 32s$, window: $2 \times 2$, output: $16 \times 16 \times 32s$]\\
5 &\textbf{CONV} [input: $16 \times 16 \times 32s$, window: $3 \times 3$, stride: $1$, kernels: $48s$, \\
&output: $16 \times 16 \times 48s$] + \textbf{BN} + \textbf{BA}\\
6 &\textbf{CONV} [input: $16 \times 16 \times 48s$, window: $3 \times 3$, stride: $1$, kernels: $64s$, \\
&output: $16 \times 16 \times 64s$] + \textbf{BN} + \textbf{BA}\\
7 &\textbf{CONV} [input: $16 \times 16 \times 64s$, window: $3 \times 3$, stride: $1$, kernels: $80s$, \\
&output: $16 \times 16 \times 80s$] + \textbf{BN} + \textbf{BA}\\
8 &\textbf{MP} [input: $16 \times 16 \times 80s$, window: $2 \times 2$, output: $8 \times 8 \times 80s$]\\
9 &\textbf{CONV} [input: $8 \times 8 \times 80s$, window: $3 \times 3$, stride: $1$, kernels: $96s$, \\
&output: $6 \times 6 \times 96s$] + \textbf{BN} + \textbf{BA}\\
10 &\textbf{CONV} [input: $6 \times 6 \times 96s$, window: $3 \times 3$, stride: $1$, kernels: $96s$, \\
&output: $4 \times 4 \times 96s$] + \textbf{BN} + \textbf{BA}\\
11 &\textbf{CONV} [input: $4 \times 4 \times 96s$, window: $3 \times 3$, stride: $1$, kernels: $128s$, \\
&output: $2 \times 2 \times 128s$] + \textbf{BN} + \textbf{BA}\\
12 &\textbf{MP} [input: $2 \times 2 \times 128s$, window: $2 \times 2$, output: $1 \times 1 \times 128s$]\\
13 &\textbf{FC} [input: $128s$, output: $10$] + \textbf{BN} + \textbf{Softmax}\\\hline

\multicolumn{2}{c}{\textbf{BC4}} \\\hline
1 &\textbf{CONV} [input: $32 \times 32 \times 3$, window: $3 \times 3$, stride: $1$, kernels: $32s$, \\
&output: $32 \times 32 \times 32s$] + \textbf{BN} + \textbf{BA}\\
2 &\textbf{CONV} [input: $32 \times 32 \times 32s$, window: $3 \times 3$, stride: $1$, kernels: $48s$, \\
&output: $32 \times 32 \times 48s$] + \textbf{BN} + \textbf{BA}\\
3 &\textbf{CONV} [input: $32 \times 32 \times 48s$, window: $3 \times 3$, stride: $1$, kernels: $64s$, \\
&output: $32 \times 32 \times 64s$] + \textbf{BN} + \textbf{BA}\\
4 &\textbf{CONV} [input: $32 \times 32 \times 64s$, window: $3 \times 3$, stride: $1$, kernels: $64s$, \\
&output: $32 \times 32 \times 64s$] + \textbf{BN} + \textbf{BA}\\
5 &\textbf{MP} [input: $32 \times 32 \times 64s$, window: $2 \times 2$, output: $16 \times 16 \times 64s$]\\
6 &\textbf{CONV} [input: $16 \times 16 \times 64s$, window: $3 \times 3$, stride: $1$, kernels: $80s$, \\
&output: $16 \times 16 \times 80s$] + \textbf{BN} + \textbf{BA}\\
7 &\textbf{CONV} [input: $16 \times 16 \times 80s$, window: $3 \times 3$, stride: $1$, kernels: $80s$, \\
&output: $16 \times 16 \times 80s$] + \textbf{BN} + \textbf{BA}\\
8 &\textbf{CONV} [input: $16 \times 16 \times 80s$, window: $3 \times 3$, stride: $1$, kernels: $80s$, \\
&output: $16 \times 16 \times 80s$] + \textbf{BN} + \textbf{BA}\\
9 &\textbf{CONV} [input: $16 \times 16 \times 80s$, window: $3 \times 3$, stride: $1$, kernels: $80s$, \\
&output: $16 \times 16 \times 80s$] + \textbf{BN} + \textbf{BA}\\
\end{tabular}
}
\end{table}
\begin{table}[h]
\centering
\resizebox{0.75\columnwidth}{!}{
\begin{tabular}{|ll|}
10 &\textbf{MP} [input: $16 \times 16 \times 80s$, window: $2 \times 2$, output: $8 \times 8 \times 80s$]\\
11 &\textbf{CONV} [input: $8 \times 8 \times 80s$, window: $3 \times 3$, stride: $1$, kernels: $128s$, \\
&output: $6 \times 6 \times 128s$] + \textbf{BN} + \textbf{BA}\\
12 &\textbf{CONV} [input: $6 \times 6 \times 128s$, window: $3 \times 3$, stride: $1$, kernels: $128s$, \\
&output: $4 \times 4 \times 128s$] + \textbf{BN} + \textbf{BA}\\
13 &\textbf{CONV} [input: $4 \times 4 \times 128s$, window: $3 \times 3$, stride: $1$, kernels: $128s$, \\
&output: $2 \times 2 \times 128s$] + \textbf{BN} + \textbf{BA}\\
14 &\textbf{MP} [input: $2 \times 2 \times 128s$, window: $2 \times 2$, output: $1 \times 1 \times 128s$]\\
15 &\textbf{FC} [input: $128s$, output: $10$] + \textbf{BN} + \textbf{Softmax}\\\hline
\multicolumn{2}{c}{\textbf{BC5}} \\\hline
1 &\textbf{CONV} [input: $32 \times 32 \times 3$, window: $3 \times 3$, stride: $1$, kernels: $32s$, \\
&output: $32 \times 32 \times 32s$] + \textbf{BN} + \textbf{BA}\\
2 &\textbf{CONV} [input: $32 \times 32 \times 32s$, window: $3 \times 3$, stride: $1$, kernels: $32s$, \\
&output: $32 \times 32 \times 32s$] + \textbf{BN} + \textbf{BA}\\
3 &\textbf{CONV} [input: $32 \times 32 \times 32s$, window: $3 \times 3$, stride: $1$, kernels: $32s$, \\
&output: $32 \times 32 \times 32s$] + \textbf{BN} + \textbf{BA}\\
4 &\textbf{CONV} [input: $32 \times 32 \times 32s$, window: $3 \times 3$, stride: $1$, kernels: $48s$, \\
&output: $32 \times 32 \times 48s$] + \textbf{BN} + \textbf{BA}\\
5 &\textbf{CONV} [input: $32 \times 32 \times 48s$, window: $3 \times 3$, stride: $1$, kernels: $48s$, \\
&output: $32 \times 32 \times 48s$] + \textbf{BN} + \textbf{BA}\\
6 &\textbf{MP} [input: $32 \times 32 \times 48s$, window: $2 \times 2$, output: $16 \times 16 \times 48s$]\\
7 &\textbf{CONV} [input: $16 \times 16 \times 48s$, window: $3 \times 3$, stride: $1$, kernels: $80s$, \\
&output: $16 \times 16 \times 80s$] + \textbf{BN} + \textbf{BA}\\
8 &\textbf{CONV} [input: $16 \times 16 \times 80s$, window: $3 \times 3$, stride: $1$, kernels: $80s$, \\
&output: $16 \times 16 \times 80s$] + \textbf{BN} + \textbf{BA}\\
9 &\textbf{CONV} [input: $16 \times 16 \times 80s$, window: $3 \times 3$, stride: $1$, kernels: $80s$, \\
&output: $16 \times 16 \times 80s$] + \textbf{BN} + \textbf{BA}\\
10 &\textbf{CONV} [input: $16 \times 16 \times 80s$, window: $3 \times 3$, stride: $1$, kernels: $80s$, \\
&output: $16 \times 16 \times 80s$] + \textbf{BN} + \textbf{BA}\\
11 &\textbf{CONV} [input: $16 \times 16 \times 80s$, window: $3 \times 3$, stride: $1$, kernels: $80s$, \\
&output: $16 \times 16 \times 80s$] + \textbf{BN} + \textbf{BA}\\
12 &\textbf{CONV} [input: $16 \times 16 \times 80s$, window: $3 \times 3$, stride: $1$, kernels: $80s$, \\
&output: $16 \times 16 \times 80s$] + \textbf{BN} + \textbf{BA}\\
13 &\textbf{MP} [input: $16 \times 16 \times 80s$, window: $2 \times 2$, output: $8 \times 8 \times 80s$]\\
14 &\textbf{CONV} [input: $8 \times 8 \times 80s$, window: $3 \times 3$, stride: $1$, kernels: $128s$, \\
&output: $8 \times 8 \times 128s$] + \textbf{BN} + \textbf{BA}\\
15 &\textbf{CONV} [input: $8 \times 8 \times 128s$, window: $3 \times 3$, stride: $1$, kernels: $128s$, \\
&output: $8 \times 8 \times 128s$] + \textbf{BN} + \textbf{BA}\\
16 &\textbf{CONV} [input: $8 \times 8 \times 128s$, window: $3 \times 3$, stride: $1$, kernels: $128s$, \\
&output: $8 \times 8 \times 128s$] + \textbf{BN} + \textbf{BA}\\
17 &\textbf{CONV} [input: $8 \times 8 \times 128s$, window: $3 \times 3$, stride: $1$, kernels: $128s$, \\
&output: $6 \times 6 \times 128s$] + \textbf{BN} + \textbf{BA}\\
18 &\textbf{CONV} [input: $6 \times 6 \times 128s$, window: $3 \times 3$, stride: $1$, kernels: $128s$, \\
&output: $4 \times 4 \times 128s$] + \textbf{BN} + \textbf{BA}\\
19 &\textbf{CONV} [input: $4 \times 4 \times 128s$, window: $3 \times 3$, stride: $1$, kernels: $128s$, \\
&output: $2 \times 2 \times 128s$] + \textbf{BN} + \textbf{BA}\\
20 &\textbf{MP} [input: $2 \times 2 \times 128s$, window: $2 \times 2$, output: $1 \times 1 \times 128s$]\\
21 &\textbf{FC} [input: $128s$, output: $10$] + \textbf{BN} + \textbf{Softmax}\\\hline
\multicolumn{2}{c}{\textbf{BC6}} \\\hline
1 &\textbf{CONV} [input: $32 \times 32 \times 3$, window: $3 \times 3$, stride: $1$, kernels: $16s$, \\
&output: $32 \times 32 \times 16s$] + \textbf{BN} + \textbf{BA}\\
2 &\textbf{CONV} [input: $32 \times 32 \times 16s$, window: $3 \times 3$, stride: $1$, kernels: $16s$, \\
&output: $32 \times 32 \times 16s$] + \textbf{BN} + \textbf{BA}\\
3 &\textbf{MP} [input: $32 \times 32 \times 16s$, window: $2 \times 2$, output: $16 \times 16 \times 16s$]\\
4 &\textbf{CONV} [input: $16 \times 16 \times 16s$, window: $3 \times 3$, stride: $1$, kernels: $32s$, \\
&output: $16 \times 16 \times 32s$] + \textbf{BN} + \textbf{BA}\\
5 &\textbf{CONV} [input: $16 \times 16 \times 32s$, window: $3 \times 3$, stride: $1$, kernels: $32s$, \\
&output: $16 \times 16 \times 32s$] + \textbf{BN} + \textbf{BA}\\
6 &\textbf{MP} [input: $16 \times 16 \times 32s$, window: $2 \times 2$, output: $8 \times 8 \times 32s$]\\
7 &\textbf{CONV} [input: $8 \times 8 \times 32s$, window: $3 \times 3$, stride: $1$, kernels: $64s$, \\
&output: $8 \times 8 \times 64s$] + \textbf{BN} + \textbf{BA}\\
8 &\textbf{CONV} [input: $8 \times 8 \times 64s$, window: $3 \times 3$, stride: $1$, kernels: $64s$, \\
&output: $8 \times 8 \times 64s$] + \textbf{BN} + \textbf{BA}\\
9 &\textbf{CONV} [input: $8 \times 8 \times 64s$, window: $3 \times 3$, stride: $1$, kernels: $64s$, \\
&output: $8 \times 8 \times 64s$] + \textbf{BN} + \textbf{BA}\\
10 &\textbf{MP} [input: $8 \times 8 \times 64s$, window: $2 \times 2$, output: $4 \times 4 \times 64s$]\\
11 &\textbf{CONV} [input: $4 \times 4 \times 64s$, window: $3 \times 3$, stride: $1$, kernels: $128s$, \\
&output: $4 \times 4 \times 128s$] + \textbf{BN} + \textbf{BA}\\
12 &\textbf{CONV} [input: $4 \times 4 \times 128s$, window: $3 \times 3$, stride: $1$, kernels: $128s$, \\
&output: $4 \times 4 \times 128s$] + \textbf{BN} + \textbf{BA}\\
13 &\textbf{CONV} [input: $4 \times 4 \times 128s$, window: $3 \times 3$, stride: $1$, kernels: $128s$, \\
&output: $4 \times 4 \times 128s$] + \textbf{BN} + \textbf{BA}\\
14 &\textbf{MP} [input: $4 \times 4 \times 128s$, window: $2 \times 2$, output: $2 \times 2 \times 128s$]\\
15 &\textbf{CONV} [input: $2 \times 2 \times 128s$, window: $3 \times 3$, stride: $1$, kernels: $128s$, \\
&output: $2 \times 2 \times 128s$] + \textbf{BN} + \textbf{BA}\\
16 &\textbf{CONV} [input: $2 \times 2 \times 128s$, window: $3 \times 3$, stride: $1$, kernels: $128s$, \\
&output: $2 \times 2 \times 128s$] + \textbf{BN} + \textbf{BA}\\
17 &\textbf{CONV} [input: $2 \times 2 \times 128s$, window: $3 \times 3$, stride: $1$, kernels: $128s$, \\
&output: $2 \times 2 \times 128s$] + \textbf{BN} + \textbf{BA}\\
18 &\textbf{MP} [input: $2 \times 2 \times 128s$, window: $2 \times 2$, output: $1 \times 1 \times 128s$]\\
19 &\textbf{FC} [input: $128s$, output: $512s$] + \textbf{BN} + \textbf{BA}\\
20 &\textbf{FC} [input: $512s$, output: $512s$] + \textbf{BN} + \textbf{BA}\\
21 &\textbf{FC} [input: $512s$, output: $10$] + \textbf{BN} + \textbf{Softmax}\\\hline

\multicolumn{2}{c}{\textbf{BH1}} \\\hline
1 &\textbf{FC} [input: $30$, output: $16$] + \textbf{BN} + \textbf{BA}\\
2 &\textbf{FC} [input: $16$, output: $16$] + \textbf{BN} + \textbf{BA}\\
3 &\textbf{FC} [input: $16$, output: $2$] + \textbf{BN} + \textbf{Softmax}\\\hline
\multicolumn{2}{c}{\textbf{BH2}} \\\hline
1 &\textbf{FC} [input: $8$, output: $20$] + \textbf{BN} + \textbf{BA}\\
2 &\textbf{FC} [input: $20$, output: $20$] + \textbf{BN} + \textbf{BA}\\
3 &\textbf{FC} [input: $20$, output: $2$] + \textbf{BN} + \textbf{Softmax}\\\hline
\multicolumn{2}{c}{\textbf{BH3}} \\\hline
1 &\textbf{FC} [input: $10$, output: $32$] + \textbf{BN} + \textbf{BA}\\
2 &\textbf{FC} [input: $32$, output: $32$] + \textbf{BN} + \textbf{BA}\\
3 &\textbf{FC} [input: $32$, output: $2$] + \textbf{BN} + \textbf{Softmax}\\\hline
\multicolumn{2}{c}{\textbf{BH4}} \\\hline
1 &\textbf{CONV} [input: $32 \times 32 \times 3$, window: $5 \times 5$, stride: $1$, kernels: $36$, \\
&output: $28 \times 28 \times 36$] + \textbf{BN} + \textbf{BA}\\
2 &\textbf{MP} [input: $28 \times 28 \times 36$, window: $2 \times 2$, output: $14 \times 14 \times 36$]\\
3 &\textbf{CONV} [input: $14 \times 14 \times 36$, window: $5 \times 5$, stride: $1$, kernels: $36$, \\
&output: $10\times 10 \times 36$] + \textbf{BN} + \textbf{BA}\\
4 &\textbf{MP} [input: $10 \times 10 \times 36$, window: $2 \times 2$, output: $5 \times 5 \times 36$]\\
5 &\textbf{FC} [input: $900$, output: $72$] + \textbf{BN} + \textbf{BA}\\
6 &\textbf{FC} [input: $72$, output: $2$] + \textbf{BN} + \textbf{Softmax}\\\hline

\end{tabular}
}
\end{table}

\newpage
\clearpage
\section{Attacks on Deep Neural Networks\vspace{-0.25cm}}\label{sec:attacks}
In this section, we review three of the most important attacks against deep neural networks that are relevant to the context of oblivious inference~\cite{tramer2016stealing,fredrikson2015model,shokri2017membership}. In all three, a client-server model is considered where the client is the adversary and attempts to learn more about the model held by the server. The client sends many inputs and receives the inference results . He then analyzes the results to infer more information about either the network parameters or the training data that has been used in the training phase of the model. We briefly review each attack and illustrate a simple defense mechanism with negligible overhead based on the suggestions provided in these works. 

\vspace{0.2em}
\noindent {\bf Model Inversion Attack~\cite{fredrikson2015model}.}
In the black-box access model of this attack (which fits the computational model of this work), an adversarial client attempts to learn about a prototypical sample of one of the classes. The client iteratively creates an input that maximizes the confidence score corresponding to the target class. Regardless of the specific training process, the attacker can learn significant information by querying the model many times. 

\vspace{0.2em}
\noindent {\bf Model Extraction Attack~\cite{tramer2016stealing}.} 
In this type of attack, an adversary's goal is to estimate the parameters of the machine learning model held by the server. For example, in a logistic regression model with $n$ parameters, the model can be extracted by querying the server $n$ times and upon receiving the confidence values, solving a system of $n$ equations. Model extraction can diminish the pay-per-prediction business model of technology companies. Moreover, it can be used as a pre-step towards the model inversion attack. 

\vspace{0.2em}
\noindent {\bf Membership Inference Attack~\cite{shokri2017membership}.}
The objective of this attack is to identify whether a given input has been used in the training phase of the model or not. This attack raises certain privacy concerns. The idea behind this attack is that the neural networks usually perform better on the data that they were trained on. Therefore, two inputs that belong to the same class, one used in the training phase and one not, will have noticeable differences in the confidence values. This behavior is called {\it overfitting}. 
The attack can be mitigated using regularization techniques that reduce the dependency of the DL model on a single training sample. 
However, overfitting is not the only contributor to this information leakage.  

\vspace{0.2em}
\noindent {\bf Defense Mechanisms.}
In the prior state-of-the-art oblivious inference solution~\cite{minionn}, it has been suggested to limit the number of queries from a specific client to limit the information leakage. However, in practice, an attacker can impersonate himself as many different clients and circumvent this defense mechanism. 
Note that all three attacks rely on the fact that along with the inference result, the server provides the confidence vector that specifies how likely the client's input belongs to each class. 
Therefore, as suggested by prior work~\cite{tramer2016stealing,fredrikson2015model,shokri2017membership}, it is recommended to augment a {\it filter} layer that (i) rounds the confidence scores or (ii) selects the index of a class that has the highest confidence score. 

\begin{enumerate}[leftmargin=*]
\vspace{-0.25cm}
\item {\it Rounding the confidence values:} 
Rounding the values simply means omitting one (or more) of the Least Significant Bit (LSB) of all of the numbers in the last layer. This operation is in fact {\it free} in GC since it means Garbler has to avoid providing the mapping for those LSBs.
\vspace{-0.25cm}
\item {\it Reporting the class label:} This operation is equivalent to computing \texttt{argmax} on the last layer. For a vector of size $c$ where each number is represented with $b$ bits, \texttt{argmax} is translated to $c\cdot(2b+1)$ many non-\texttt{XOR} (\texttt{AND}) gates. 
For example, in a typical architecture for MNIST (e.g., BM3) or CIFAR-10 dataset (e.g., BC1), the overhead is 1.68E-2\% and 1.36E-4\%, respectively.
\end{enumerate}
\vspace{-0.25cm}
Note that the two aforementioned defense mechanisms can be augmented to any framework that supports non-linear functionalities~\cite{minionn,chameleon,deepsecure}. However, we want to emphasize that compared to mixed-protocol solutions, this means that another round of communication is usually needed to support the filter layer. Whereas, in \sys{} the filter layer does not increase the number of rounds and has negligible overhead compared to the overall protocol.

\end{document}